# Optical design for CETUS: a wide-field 1.5-m aperture UV payload being studied for a NASA probe class mission study

Robert A. Woodruff
William C. Danchi
Sara R. Heap
Tony Hull
Stephen E. Kendrick
Lloyd R. Purves
Michael S. Rhee
Eric Mentzell
Brian Fleming
Marty Valente
James Burge
Ben Lewis
Kelly Dodson
Greg Mehle
Matt Tomic

# Optical design for CETUS: a wide-field 1.5-m aperture UV payload being studied for a NASA probe class mission study

**Robert A. Woodruff,[a,*] William C. Danchi,[b] Sara R. Heap,[b] Tony Hull,[c,d] Stephen E. Kendrick,[c] Lloyd R. Purves,[b] Michael S. Rhee,[b] Eric Mentzell,[b] Brian Fleming,[e] Marty Valente,[f] James Burge,[f] Ben Lewis,[f] Kelly Dodson,[g] Greg Mehle,[g] and Matt Tomic[g]**
[a]Woodruff Consulting, Boulder, Colorado, United States
[b]NASA Goddard Space Flight Center, Greenbelt, Maryland, United States
[c]Kendrick Aerospace Consulting LLC, Lafayette, Colorado, United States
[d]University of New Mexico, Albuquerque, New Mexico, United States
[e]Laboratory for Atmospheric and Space Physics, Boulder, Colorado, United States
[f]Arizona Optical Systems, Tucson, Arizona, United States
[g]Northrop Grumman Innovation Systems, San Diego, California, United States

**Abstract.** As part of a study funded by NASA headquarters, we are developing a probe-class mission concept called the Cosmic Evolution through UV Spectroscopy (CETUS). CETUS includes a 1.5-m aperture diameter telescope with a large field of view (FOV). CETUS includes three scientific instruments: a far ultraviolet (FUV) and near ultraviolet (NUV) imaging camera (CAM); a NUV multiobject spectrograph (MOS); and a dual-channel point/slit spectrograph (PSS) in the Lyman ultraviolet (LUV), FUV, and NUV spectral regions. The large FOV three-mirror anastigmatic (TMA) optical telescope assembly (OTA) simultaneously feeds the three separate scientific instruments. That is, the instruments view separate portions of the TMA image plane, enabling parallel operation by the three instruments. The field viewed by the MOS, whose design is based on an Offner-type spectrographic configuration to provide wide FOV correction, is actively configured to select and isolate numerous field sources using a next-generation micro-shutter array. The two-channel CAM design is also based on an Offner-like configuration. The PSS performs high spectral resolution spectroscopy on unresolved objects over the NUV region with spectral resolving power, $R \sim 40{,}000$, in an echelle mode. The PSS also performs long-slit imaging spectroscopy at $R \sim 20{,}000$ in the LUV and FUV spectral regions with two aberration-corrected, blazed, holographic gratings used in a Rowland-like configuration. The optical system also includes two fine guidance sensors, and wavefront sensors that sample numerous locations over the full OTA FOV. In-flight wavelength calibration is performed by a wavelength calibration system, and flat-fielding is also performed, both using in-flight calibration sources. We describe the current optical design of CETUS and the major trade studies leading to the design.  

## 1 Introduction

The planned termination of the highly successful Hubble Space Telescope (HST) mission will eliminate scientific access to large aperture ultraviolet (UV)-optimized space-borne instrumentation in the reasonably near future. The Cosmic Evolution through UV Spectroscopy (CETUS) mission concept, if approved and implemented, will partially fill this void.[1,2] CETUS is a NASA headquarters-selected probe-class mission concept which includes a 1.5-m aperture diameter large field of view (FOV) optical telescope assembly (OTA) optimized for imaging and spectroscopy with three science instruments (SIs) that observe within the 100- to 400-nm spectral region. The CETUS design enables simultaneous far UV/near UV (FUV/NUV) wide-field imaging, NUV multiobject spectroscopy, and LUV/FUV long-slit spectroscopy and NUV point source spectroscopy. The CETUS mission concept design started in earnest in the spring of 2017 and the entire study will be completed by the end of calendar year 2018. This paper describes the optical design resulting from this study.

*Address all correspondence to Robert A. Woodruff, E-mail: raw@colorado.edu

## 2 Design Overview

### 2.1 *Science Goals Determine CETUS Design*

The CETUS observatory is designed to study galaxy evolution, providing long exposure observations of a large number of objects. The science themes include Cosmic Evolution of Galaxies, Stars, and Exo-Planets, The Modern Universe: What It's Made of; How It Works, Transients, and Deep Surveys.[1,2] To fit within a \$1 billion budget, a moderate size of 1.5-m diameter OTA aperture size is proposed. Observations of a large sample of objects are provided by multipixel detectors and parallel observing. The CETUS observatory will be in a Sun/Earth L2 large halo orbit which enables continuous, uninterrupted viewing of stellar objects. A typical observation with the multiobject spectrograph (MOS) and the camera (CAM) will be ~10 h long with the CCD detectors readout every 30 min, in order to reduce the effect of cosmic ray/charged particle detections. In this orbit,

CETUS is about 0.01 AU ($\sim 1.5 \times 10^6$ km) on the anti-Sun of the Earth. The Sun, the Earth, and the Moon are each always on the sun side of CETUS and thus light from them will be excluded from entering the OTA by its sunshield. With an 85-deg solar exclusion angle, a full $2\pi$ steradian anti-Sun hemisphere of the universe is available for viewing at any given time. The full $4\pi$ steradian view of space is accessible over a 1-year time frame because of the Earth's orbital progression around the Sun.[3,4]

CETUS observes and records spectrally filtered images, as well as low- and high-resolving power spectroscopy of galaxies, in the vacuum UV spectral region. Highly efficient photon detection design is enabled by (1) using an all-reflective optical design of the OTA and the instruments, (2) using Lyman-$\alpha$-optimized [or as an option predicated on successful technology development, Lyman UV (LUV)-optimized protected Al + LiF coating] reflective mirror coatings, (3) minimizing the number of reflections in each optical path, (4) using field-sharing that permits simultaneous observation by SIs, and (5) using detectors with minimum number of windows and high quantum efficiency.

Table 1 summarizes the top-level requirements of the OTA and SIs.

The SIs are as follows:

- A NUV MOS with field objects selected for observation by a next-generation microshutter array (MSA). The MOS provides 1-pixel spectral resolving power of $R \sim 1000$ over 180- to 350-nm wavelength range.
- A CAM covering with two spectral channels: FUV (115 to 195 nm) and NUV (175 to 400 nm).
- A two-mode point/slit spectrograph (PSS) covering the LUV, FUV, and NUV spectral regions. The NUV PSS echelle spectrograph obtains 2.5-pixel resolving power of $R \sim 40{,}000$ over wavelengths 178 to 354 nm. The LUV/FUV PSS provides 1-resolution element (resel) spectral resolving power of $R \sim 20{,}000$ over wavelengths 100 to 180 nm. The PSS modes share a single fixed entrance slit located at the TMA focus centered in the OTA field. The NUV mode observes isolated "point sources." The LUV/FUV mode is used for imaging spectroscopy over a 6-arc min-long slit using Rowland-like aberration corrected, holographic gratings based on Cosmic Origins Spectrograph on HST (COS-HST). This is a true long-slit imaging spectroscopy capability.

The NUV channels of the CAM and PSS and MOS, each use a $4\text{k} \times 4\text{k}$ e2V Euclid CCD273-84 CCD detector with 12-$\mu$m pixels. The FUV CAM uses a sealed CsI microchannel plate (MCP) detector with 20-$\mu$m resels on a $50\ \text{mm} \times 50\ \text{mm}$ flat photocathode, The LUV/FUV PSS uses an open CsI MCP detector with 22-$\mu$m resels on a $70\ \text{mm} \times 200\ \text{mm}$ curved photocathode.

Each instrument samples a dedicated separate portion of the large well-corrected OTA FOV. This approach to field-sharing enables simultaneous science observations by all instruments, greatly contributing to the high observational efficiency of CETUS. The resultant FOV shown in Fig. 1 illustrates this implemented approach. The overall OTA 0.65 deg × 1.1 deg FOV is sampled by each SI, fine guidance sensor (FGS), and wavefront sensor (WFS) channel as shown. These field-limiting features are implemented in the TMA focal assembly (TFA) which resides at the TMA focus. The OTA forms sharp images at the TMA focus of objects in the OTA FOV. The light then diverges into the separate SIs, namely, the MOS, the CAM, and the PSS.

The TFA also includes the ends of five single-mode optical fibers, each conjugated to a WFS on autocollimation of the OTA. During ground system optical alignment, light from each fiber projects a full aperture beam through the OTA that exits as a collimated beam to a full aperture flat mirror which, in turn, reflects the return beam filling the OTA entrance pupil. The OTA reimages each fiber at the TMA focus. The WFSs can sample the return image to verify image quality. The return beam field image, when directed to the CAM or the MOS field locations via small tilt of the autocollimating flat, will form field images at the respective flight NUV detector. The fiber-transmitted wavelengths, being >320 nm, lie in the spectrally sensitive band of each of these NUV instruments and enable preflight full system test of the OTA + SI optical alignment (see Sec. 4, for more details).

**Table 1** Top-level requirements of the OTA and SIs.

| | | | Spatial | | | | Spectral | | |
|---|---|---|---|---|---|---|---|---|---|
| | | | FOV (deg) | FWHM (arc sec) | | | Resolving power, $R = \lambda/\Delta\lambda$ | | |
| Instrument | $\lambda$ min (nm) | $\lambda$ max (nm) | $(X, Y)$ | | Over FOV | @ $\lambda$ (nm) | $\lambda$ min | $\lambda$ max | |
| OTA @ TMA focus | 115 | 1100 | (1.10, 0.652) | 0.083 | 0.237 | 200 | NA | NA | |
| OTA + FUV CAM | 115 | 195 | (0.290, 0.290) | 0.059 | 0.182 | 140 | NA | NA | |
| OTA + NUV CAM | 175 | 400 | (0.290, 0.290) | 0.067 | 0.177 | 215 | NA | NA | |
| OTA + MOS | 180 | 350 | (0.290, 0.290) | 0.055 | 0.153 | 250 | 730.0 | 1420.0 | 1 pixel |
| | (nm) | (nm) | Slit (arc min) | | FWHM (arc sec) | @ $\lambda$ (nm) | Resolving Power, $R = \lambda/\Delta\lambda$ (Over spectra) | | |
| OTA + NUV echelle PSS | 178 | 354 | 6 | | 0.278 | 271 | $3.92 \times 10^{04}$ | $4.14 \times 10^{04}$ | 2.5 pixels |
| OTA + LUV/FUV Rowland PSS | (100) 115 | 180 | | | 0.260 | | 20,000 | | 1 pixel |

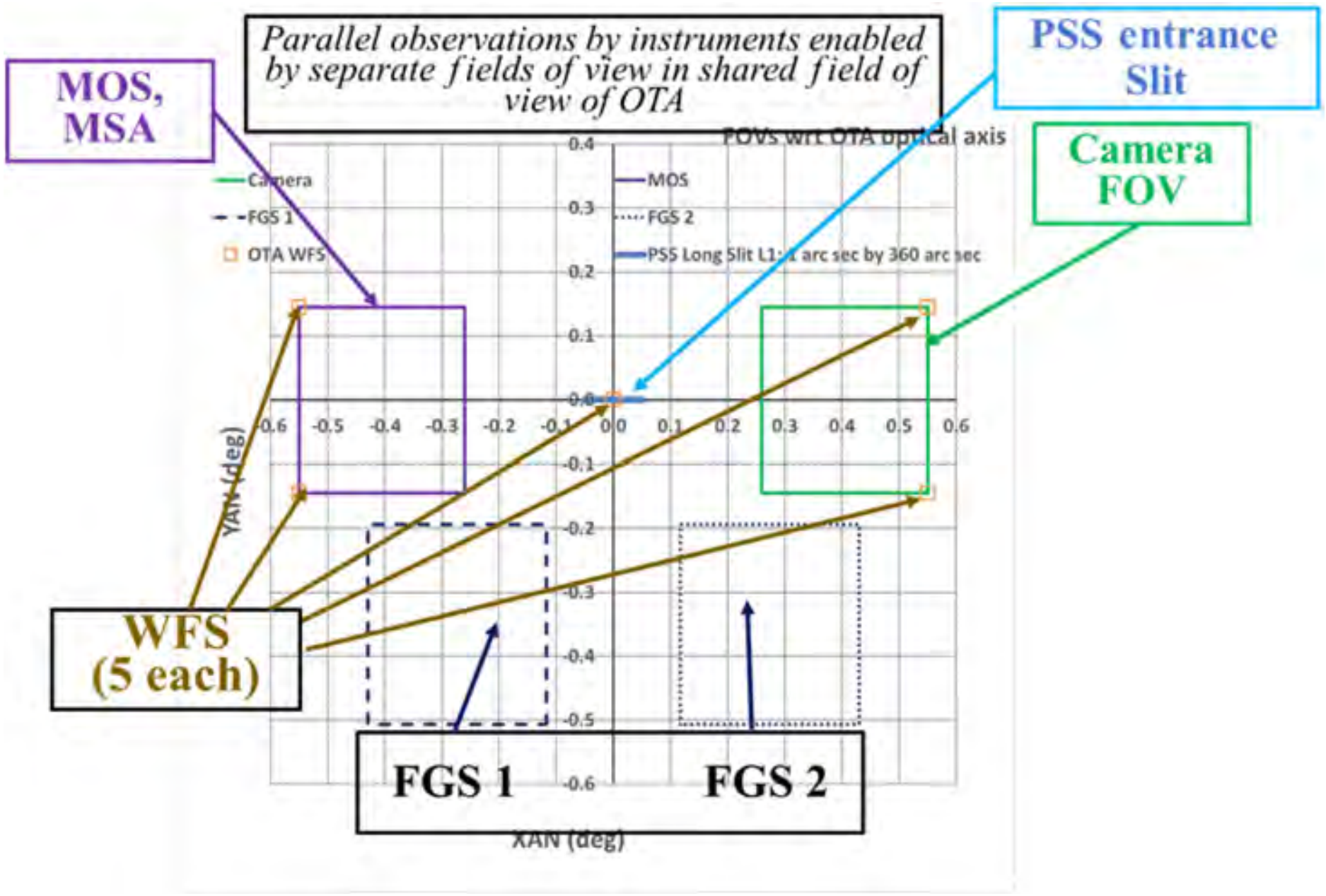

**Fig. 1** Shared-splitting of CETUS FOV. Fiber optics alignment sources (not shown) also reside in this TFA. Individual fields as imaged onto the sky by the OTA.

Over the course of the observatory's lifetime, the OTA and SIs must maintain optical alignment, including focus. Based on the image error budget, discussed in Sec. 3, the design allocates 170 $\mu$m to defocus at the TMA focus. This corresponds to the 5-$\mu$m despace at the secondary mirror (SM). OTA optical alignment is achieved using five WFSs near the four corners and center of the OTA FOV, which provide error signals to realign the OTA SM to correct the TMA image of the stellar field. This method allows compensation and correction of focus, third-order coma, image location, and image plane tip/tilt of the image. The SIs maintain their individual internal optical alignment by the design using low-thermal expansion stable materials for structure and optics, stable designs for the mirror mounts, and temperature control. Their location in the facility is well isolated from external thermal environmental extremes, so their temperatures can readily be maintained stable and uniform. The M2 optic of both the MOS and the CAM is mounted on a tip/tilt/focus mechanism. Thus, their internal alignment can be adjusted to compensate internal misalignments. Each WFS is metered to the OTA structure, providing true sensing of OTA wavefront errors.

As shown in Figs. 2 and 3, fixed fold mirrors near the aberrated Cassegrain focus of the OTA, where the field is imaged by the first two mirrors of the TMA, direct objects in the indicated FOVs to the FGSs sensors. The image at each FGS detector is formed by refractive lenses that reimage the field imaged by the primary/secondary mirror Cassegrain pair onto the respective FGS detector. These lenses correct the image quality to a level suitable for precise star centroid detection. These provide error signals to the attitude control system (ACS) for line-of-sight fine body pointing.

When wavefront sensing is performed, light to the SIs is blocked by a pupil insert mirror (PIM) that rotates into the region where the OTA real exit pupil lies. This PIM reflects the OTA field to five WFS detectors which provide error signals to adjust the SM to collimate the TMA. This sensing does not interfere with the view of guide starlight sensing by the FGS, thus pointing control is not interrupted.

When the wavelength calibration of the PSS is performed, light to the SIs is blocked by a calibration insert mirror (CIM) that rotates into the region near Cassegrain focus. This mirror reflects the wavelength calibration source within the wavelength calibration system (WCS) to overfill the PSS entrance slit, providing known spectral lines for wavelength calibration. Again, the FGS is not affected and fine pointing control is not interrupted.

A key major trade in the overall architecture involves placement of the auxiliary systems: FGS, WFS, and WCS. We considered picking off the beam to the FGS and inserting the beam from the WCS at the TMA focus. This is not feasible because the SIs fully fill this region. Instead we utilize the region near Cassegrain focus for the FGS and the WCS. The WFS must sense the aberrated wavefront of the whole TMA, so it must reside in conjugate to the TMA focus. The OTA exit pupil provides a convenient location for the pick-off mirror to direct field stars to the multiple WFS detectors.

The design specifically allows integration of each instrument without affecting the alignment of previously installed instruments or of the OTA. This feature reduces the programmatic and schedule risks during integration and test. Problems with any instrument will not affect the integration of other instruments or of the OTA. Programmatic risk is also reduced by this approach, thus allowing procurement of OTA and SIs from numerous outside sources with interfaces controlled by the interface control documents (ICDs).

To minimize the number of reflections and reflection loss, we use no fold mirrors in the OTA. The result is a long OTA

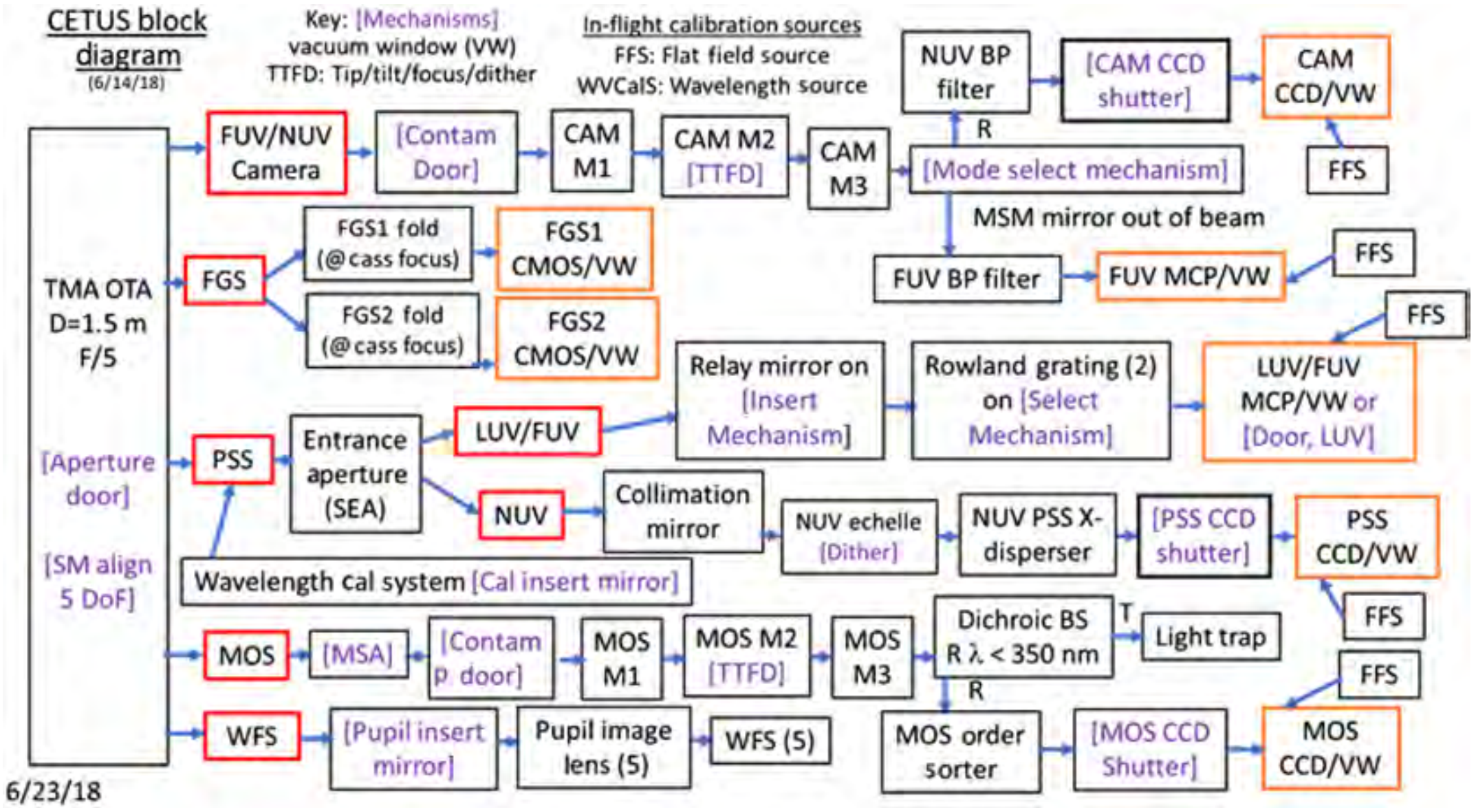

**Fig. 2** Optical system block diagram showing interrelationship of major subsystems and science capabilities.

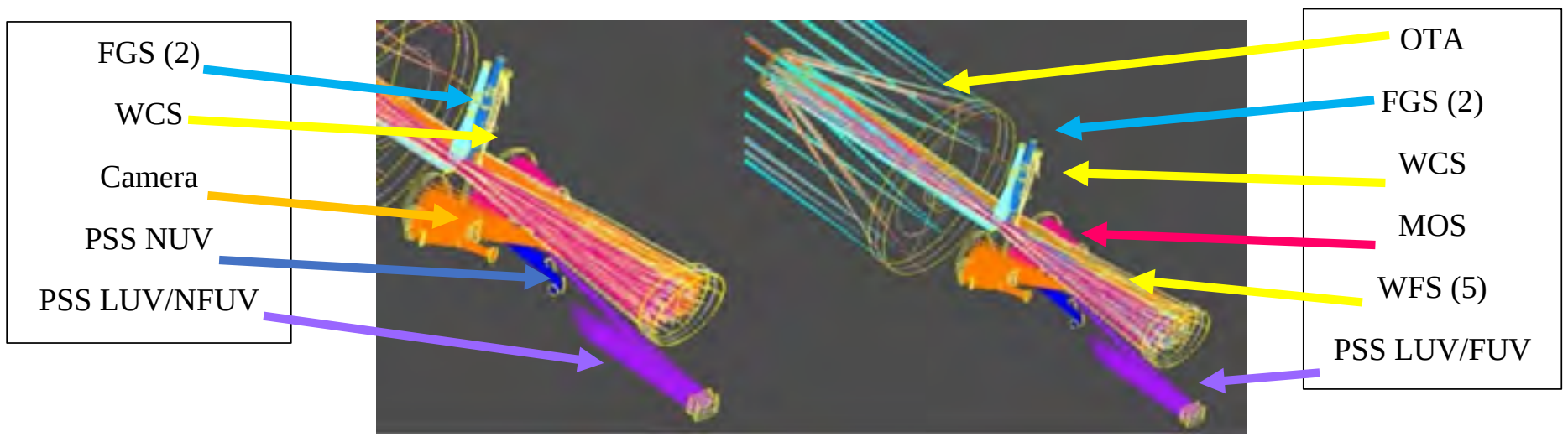

**Fig. 3** CETUS OTA + instruments.

(~4.3 m from SM to TM). Our baseline launch vehicle, a Falcon 9, easily accommodates this length.

## 2.2 *Optical Telescope Assembly*

The OTA design form is an all-reflective three-mirror anastigmatic (TMA) telescope with a flat field. A two-mirror OTA would not achieve the required image quality over the required FOV. A three-mirror solution provides the needed corrected FOV using a minimum number of reflections. A TMA provides sufficient design variables to correct the following third-order aberrations: spherical aberration, coma, and astigmatism. Its physical parameters when properly selected yield a flat field image surface (i.e., zero Petzval curvature).[5–7]

The corrected FOV of the f/5, focal length of 7.5-m, OTA is ~$1.1 \times 0.65$ deg. The 1.5-m diameter OTA entrance pupil/aperture stop is at the front surface of the $f/1.448$ PM. The OTA has a centrally obscured on-axis aperture. The three conic mirrors of the OTA share a common optical axis. To physically separate the optical light beam reflected by the tertiary mirror (TM) from the beam incident on the TM, the OTA is off-axis in field by 0.45 deg relative to its optical axis, thereby providing physical access to the TMA focus region and the real exit pupil of the OTA. The radii and conic constant (CC) of the OTA mirrors are (where CC = 0 is a spherical surface and CC = −1.0 is a paraboloid of revolution):

- **PM:** Concave ellipsoid of revolution, $R = -4{,}343.26384$ mm with CC = −0.9272559, vertex at global $Z = 0$.
- **SM:** Convex, $R = -1{,}031.38354$ mm with magnification $m_2 = +5.7426$, hyperboloid of revolution with CC = −1.4227702, vertex at global $Z = -1{,}745.74061$ mm.
- **TM:** Concave, $R = -1{,}352.57631$ mm with magnification $m_3 = -0.6014$, ellipsoid of revolution with CC = −0.5337356, vertex at global $Z = +2{,}500.81119$ mm.
- **Back focal distance of PM/SM pair:** Axial distance from PM vertex to Cassegrain focus by the PM/SM pair, $b = 700.0$ mm.

The large Cassegrain back focal distance provides depth that enables designing of a stiff stable member main structural assembly on which the PM, the SM support truss, the TM support truss, and the three instruments together with the WFS, FGS, and WCS are supported.

Figure 4(a) displays the OTA design from a side view, whereas Fig. 4(b) illustrates the OTA design with the red rays ending at the field image, i.e., the TMA focus plane, and the blue

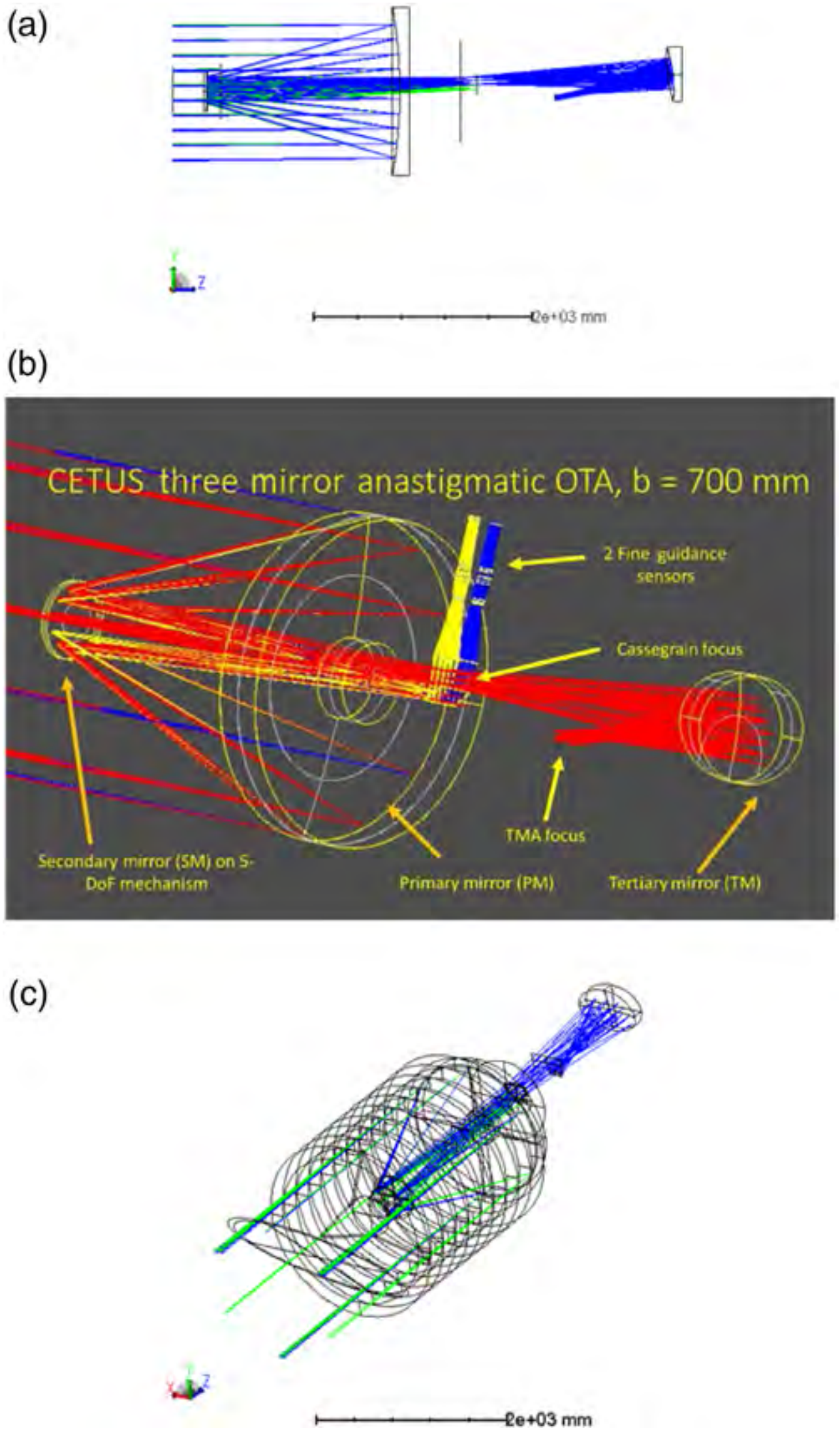

**Fig. 4** (a) OTA with (b) FGS, and with (c) baffles.

and yellow rays reflected by pick-off mirrors near Cassegrain focus to the two FGSs. The TFA, composed of the MOS MSA, a field-limiting square aperture for the CAM, the PSS spectral entrance slit, and fixed termination ends of fly-away single-mode fibers, lies at the TMA focus. The OTA and its stray light baffles are shown in Fig. 4(c). The SM is supported by six struts that extend from the PM housing to the SM mount. The OTA is baffled to off-axis sources with a main baffle, surrounding the 1.5-m light beam extending from the PM to beyond the SM, and with classical Cassegrain central baffles: the SM cone baffle and the PM central baffle. In addition, the sunshield assembly surrounds the main baffle and extends from behind the PM to well beyond the SM. Its outward end is angled at 45 deg to exclude sunlight from entering the OTA even at solar exclusion angle of 85 deg. A reusable door, mounted on the tilted end of the sunshield, is open for science operations and closed to protect from imaging the sun during safing and to protect from contamination during orbital maintenance propellant firings and during launch. This door also provides prelaunch contamination control as do other design features, including preflight $N_2$ purge, maintaining optics temperatures warmer than other coldest surfaces, screening of materials for low outgassing, and other system controls.

## 2.3 *Multiobject Spectrograph*

Figure 5(a) shows side and top views of the design of the combined OTA plus MOS, whereas Fig. 5(b) shows the design rays and features of the MOS. The MOS imaging spectrometer observes numerous stellar objects simultaneously recording the spectral content of each object at 1-pixel $R \sim 1000$ over the spectral range from 180 to 350 nm. The $f/5.24$ MOS reimages a $1045 \times 1045$ arc sec portion of the OTA FOV via a three-mirror, all-reflective, Offner-like, nearly one-to-one, imaging spectrometer.[8–10] The three-mirrors are nominally concentric with M1 and M3 concave radius of curvature of 800 mm and M2 convex with radius of curvature of 400 mm. The M2 mirror is a convex, spherical, reflective diffraction grating with 140 grooves per millimeter blazed at 250 nm. Mirrors M1 and M3 are high-order aspheric mirrors. To reduce fabrication complexity, the design maintains the MOS M2 as a spherical diffraction grating.

Simultaneous spectral imaging of numerous objects is enabled by the configurable MSA located at the sharp image provided by the OTA at the TMA focal plane. The 38 mm $\times$ 38 mm MSA has $380 \times 190$ individually selectable rectangular shutters. Because each shutter can be independently controlled, any combination of shutters can be commanded to open or close. Each shutter is 100-$\mu$m wide (2.75 arc sec) in spectral dispersion direction and 200-$\mu$m wide (5.50 arc sec) in the cross-dispersion direction. When observing, opened apertures match the actual image locations at the TMA focus of stellar field objects. The field locations are selected to eliminate spectrally dispersed overlap of the spectrum from nearby objects.

A 4 K $\times$ 4 K, e2V Euclid CCD273-84 with 12-$\mu$m pixels detects the spectrum of every selected object, thereby providing time-efficient parallel observations of the galaxy field. The MOS CCD housing is a vacuum enclosure with an 8-mm-thick $MgF_2$ vacuum window that maintains vacuum integrity. Placed between the M3 mirror and the CCD is a 2-mm-thick UV-grade fused silica order sorter that absorbs light of $\lambda < 160$ nm, effectively blocking second- and higher-order light. One option studied a fused silica window on the CCD housing for this function. Unfortunately, the image quality and throughput at the 180-nm end of the spectrum suffered from dispersion and absorption by the thick, fused silica substrate required for vacuum integrity. Therefore, a lower dispersion material is selected for the window ($MgF_2$) and a thinner order sorter is selected that is more optimal for its order-sorting function. Note that the CAM CCD package also uses a $MgF_2$ window to minimize image quality degradation due to dispersion quality, and so a similar design of the CCD assembly is enabled by these design choices.

The CCD light-sensitive surface is cooled to $162 \pm 5$ K to reduce noise due to dark current, while maintaining the window at room temperature. The vacuum housing permits ground testing and alignment in a 1-atm environment without adding to the cost of testing that would be required if ground testing required a vacuum chamber. While in-flight, the accumulation of contamination on the window surface is avoided by maintaining the window at room temperature. Periodically the window temperature is increased slightly by heaters adjacent to the window to boil off accumulated contaminates.

The optical design of the MOS and CAM is based on a modified Offner, concentric, three-mirror relay design, at unity magnification. A true Offner relay uses three concentric spherical mirrors for a fully symmetric optical relay with unity

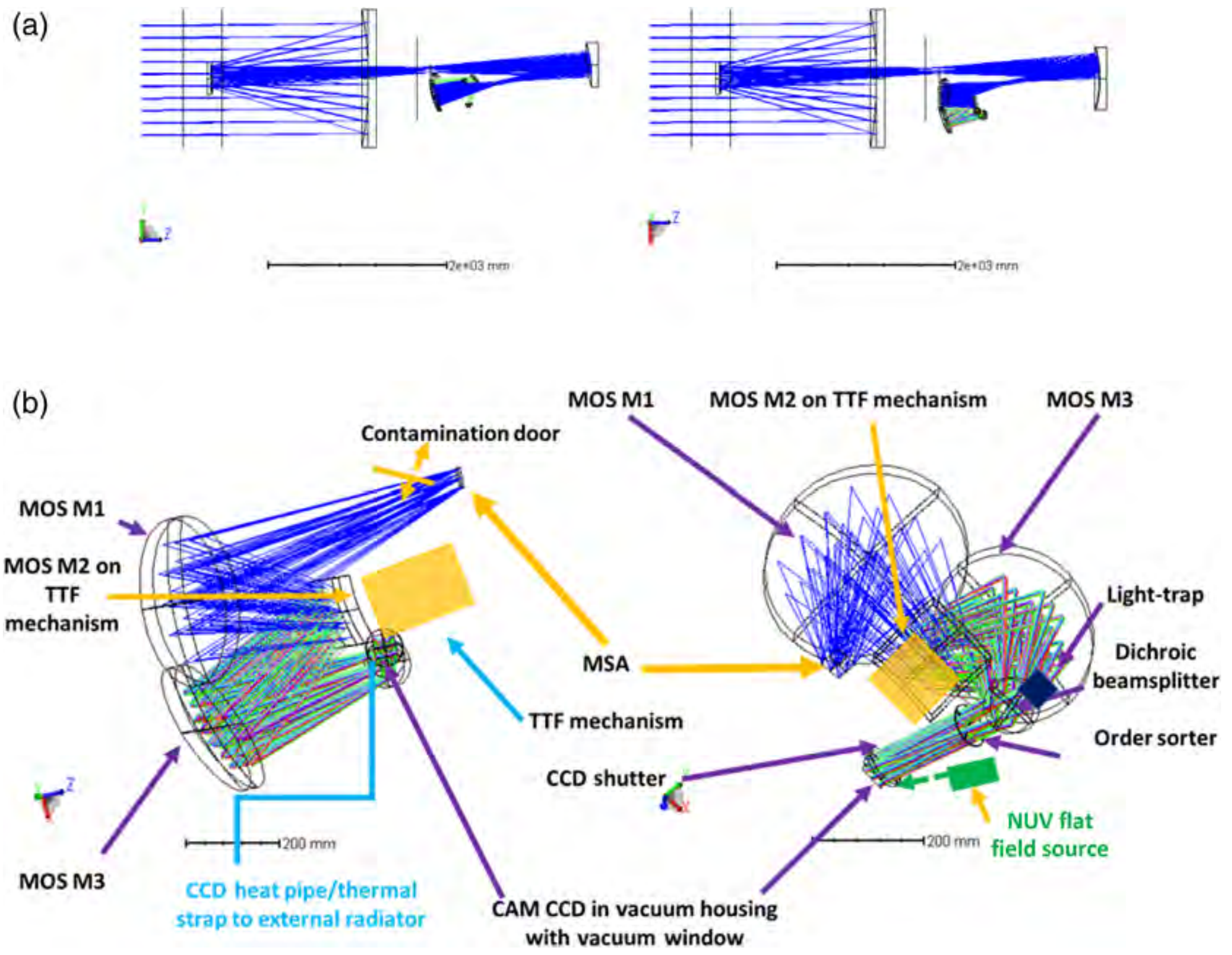

**Fig. 5** (a) OTA + MOS (left shows the side view and right shows the top view). (b) MOS: Mirrors, mechanisms, dichroic beamsplitter, flat-field calibration sources, and FPA interface.

magnification. In this ideal configuration, the radii of curvature of concave mirror, M1 and M3, are identical, and the radii of curvature of the convex mirror M2 are one-half this parameter. (In the MOS design, the M2 mirror is a reflective grating. In the CAM, M2 is a convex mirror.) To maintain the essential symmetry in a true Offner, the aperture stop is at the surface of M2.[11] If the stop of the system cannot be at M2, the ideal symmetry of the relay is compromised.

For CETUS, the optical design of both the MOS and the CAM encounters and addresses this issue by modifying the Offner design. For a true Offner, mirror M2 lies one focal length distance from mirror M1. Thus, the exit pupil of an optical system feeding a true Offner must lie at infinity to properly position the pupil at mirror M2. In CETUS, the OTA forms an image of the field near the plane of the shared centers of curvature of the three mirrors and forms a real exit pupil at a finite distance before the TMA focal surface (~290 mm). Thus, unless we include additional optical elements, such as a field lens or mirror, the pupil cannot lie at the desired M2 mirror location.

Our design for the MOS and CAM must be a compromise, as no clean image of the pupil near M2 is possible. We still achieve acceptable optical correction and performance. The solution includes use of highly aspheric mirror surfaces in the Zemax optimization.

The M2 mirror is supported on a tip/tilt/focus mechanism. The MOS can be focused independently of the OTA and the CAM. Tip/tilt adjustment enables dithering the image at the FPA to sense detector pixel sensitivity variations.

A dichroic beamsplitter following the M3 mirror will transmit visible light to the light trap to reduce red leak in the spectrograph spectrum. The effective resel width of the 12-$\mu$m FPA pixel imaged at the TMA focal surface by the MOS optics is 396 mas in field.

### 2.4 *FUV and NUV Camera*

Figure 6(a) shows the design of the combined OTA plus CAM, whereas Fig. 6(b) shows a detailed design of CAM. The FUV and NUV CAMs form images of numerous celestial objects simultaneously in their respective spectral regions. The $f/5$ CAM reimages a $1045 \times 1045$ arc sec portion of the OTA FOV via a three-mirror, all-reflective, Offner-like, one-to-one imager, as discussed in Sec. 2.3. The three mirrors are nominally concentric with M1 concave radius of curvature of 776.985 mm, M3 concave radius of curvature of 657.935 mm and M2 convex radius of curvature of 409.949 mm. All three mirrors are high-order aspheric mirrors.

Wide field imaging is enabled by the CAM mirrors relaying the sharp image provided by the OTA at the TMA focal plane to the two CAM detectors, respectively. The FUV and NUV bands share the three powered mirrors. The band is selected by a plano fold mirror on a mode select mechanism which lies between mirror M3 and the detectors. To preserve FUV photons relative to NUV photons, the FUV mode is selected by removing this fold mirror from the optical path. When inserted, the fold mirror reflects the field to the NUV detector.

Each path has selectable spectral bandpass filters on a filter wheel between the fold mirror and the detector. Based on the experience from the HST instruments, such as the solar blind channels in the HST Space Telescope Imaging Spectrograph (STIS-HST) and the Advanced Camera for

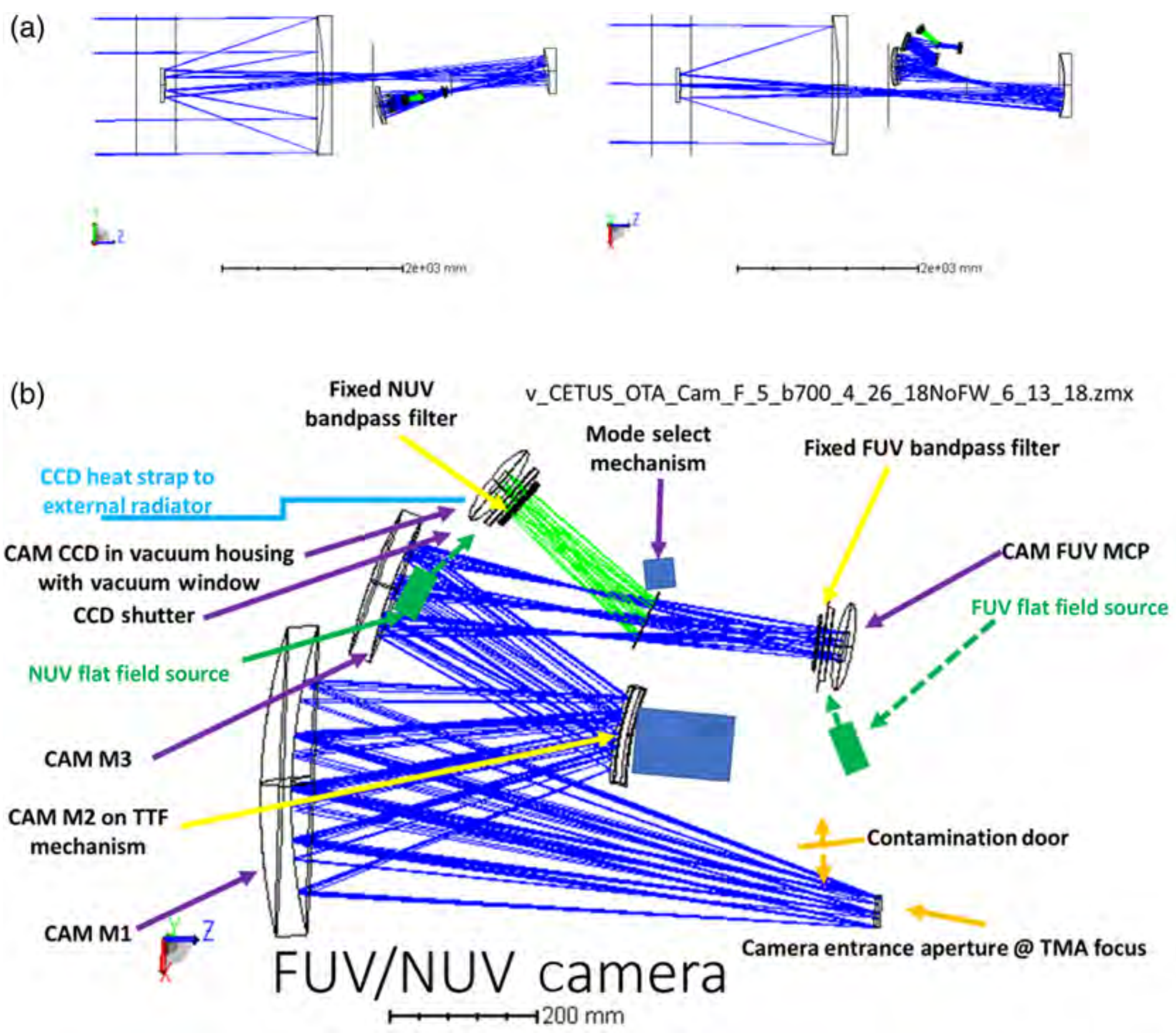

**Fig. 6** (a) OTA + CAM (left shows side view and right shows top view). (b) CAM: Mirrors, mechanisms, flat-field calibration sources, and FPA interface.

Surveys (ACS-HST), the FUV filter is a single-element uncoated crystal. The NUV filters are a dual-element bandpass filters. Following the experience from the design of ACS-HST instrument, the NUV filters are air-spaced (i.e., vented) dual-element filters with coatings on each of the four plano surfaces. These define the in-band spectral throughput and block the out-of-band signal from the bandpass upper wavelength to the red-cutoff of the CCD and from the bandpass lower wavelength to the blue cutoff of the CCD.[12] One substrate is UV-grade fused silica and the other is fused silica or colored glass.

The image in the FUV mode is detected by a 2 K × 2 K sealed CsI solar blind MCP detector with 20-$\mu$m effective resels. The MCP housing is a vacuum enclosure with a $MgF_2$ window. The detector operates at room temperature. One 20-$\mu$m resel, imaged by the PSS internal optics relay to the TMA focal surface, subtends an angle 550 mas in field.

The image in the NUV mode is sensed by a 4 K × 4 K, e2V Euclid CCD273-84 with 12-$\mu$m pixels, identical to the MOS CCD. The CAM CCD housing is a vacuum enclosure with a $MgF_2$ window. The benefits of this design choice for the integration and test effort and the on-orbit operation were previously discussed in Sec. 2.3.

As in the MOS, the M2 mirror is supported on a tip/tilt/focus mechanism, so the CAM can be focused independently of the OTA and the MOS. Tip/tilt adjustment enables dithering the image at the FPA to correct the detector sensitivity variations.

The effective resel width of the 12-$\mu$m FPA pixel imaged at the TMA focal surface by the CAM optics is 330 mas in field.

### 2.5 *LUV/FUV/NUV Point/Slit Spectrograph*

Figure 7(a) displays the design of the combined OTA plus PSS, whereas Figs. 7(b) and 7(c) show the detailed design of the PSS. The LUV/FUV Rowland-like spectrograph and the NUV echelle spectrograph together comprise the CETUS PSS. The PSS records high resolving power spectra of individual sources that the OTA images onto the shared entrance slit at the TMA focus. The OTA light passes through the entrance slit and then diverges at $f/5$ to one of the two mirrors: a fixed NUV parabolic collimation mirror or a selectable LUV/FUV relay mirror. The NUV echelle design is similar to the echelle mode designs of HST instruments: the Goddard high-resolution spectrograph (GHRS-HST) and the STIS-HST. The LUV/FUV design has heritage from the COS-HST, utilizing only a single optic, the disperser, for high optical efficiency spectroscopy.

To select the $R \sim 40{,}000$ NUV mode, the LUV/FUV relay mirror is withdrawn from the OTA $f/5$ postentrance slit diverging beam. The NUV parabolic mirror collimates the diverging beam, reflecting it to a plano NUV echelle diffraction grating, which has 100 grooves/mm, $\theta = 42.0$-deg blaze angle, and $\tan(\theta) = 0.90$. CETUS uses it in diffraction orders from $m = 38$ through 74. This echelle grating reflects the dispersed beam to a NUV parabolic cross disperser, used in first order, with focal length of

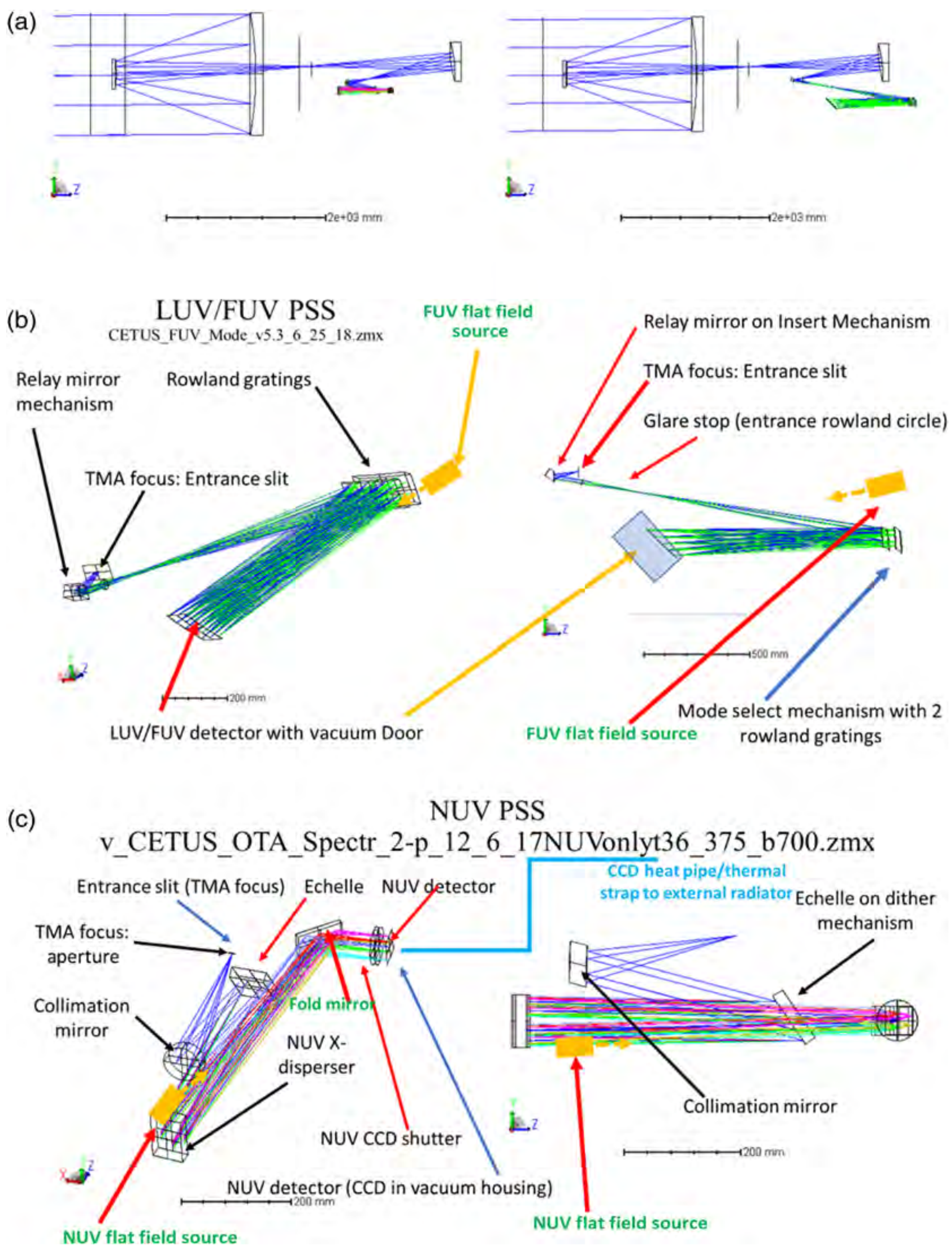

**Fig. 7** (a) OTA + LUV/FUV/NUV PSS (left panel shows NUV and right panel shows LUV/FUV). (b) LUV/FUV PSS. (c) NUV PSS

750 mm, 333.33 grooves/mm, blaze wavelength of 265.0 nm, blaze angle of 2.53 deg, and Wadsworth angle of 5.07 deg. The NUV spectrograph is a Czerny–Turner configuration. The cross disperser uses a Wadsworth aplanatic configuration, with an angle of diffraction near zero at wavelength band center. The resulting third-order spherical aberration and third order are zero at the band-center focus position.

The resultant $f/15$ beam from the cross-disperser is imaged onto the NUV 4 K $\times$ 4 K, e2V Euclid CCD273-84 with 12-$\mu$m pixels, creating an echellogram. (This is the same type of detector used for the MOS and NUV CAM.) This mode achieves the spectral resolving power, $R \sim 40{,}000$, over the spectral region from 178.04 to 353.67 nm, with 2.5-pixel effective pixel width. The CCD is housed in a vacuum enclosure with UV-grade fused silica window. The window provides order sorting that absorbs light of $\lambda < 160$ nm, effectively blocking second- and higher-order light.

The effective resel width of the 12-$\mu$m FPA pixel imaged at the TMA focal surface by the NUV optics is 349 mas in field. As described in Sec. 3.2, the chosen 2.5-pixel resel width, instead of a 1-pixel width, relaxes the requirements on the OTA image quality.

When using either of the two LUV/FUV modes (G120 or G150), a mechanism inserts the LUV/FUV relay mirror into the diverging beam from the OTA, thereby blocking light into the NUV PSS. This relay mirror reimages the OTA $f/5$ beam that diverges from the entrance slit to an $f/17.5$ converging beam, thus creating a partially corrected field image at the location of an oversized entrance "slit" glare stop on a pseudo-Rowland circle. The resultant beam then diverges to one of the two holographic, aberration-corrected, ion-etched blazed "Rowland-like" gratings. One resel spectral resolving power of 20,000 is achieved. The image quality along the entrance slit length is sufficient to allow imaging spectroscopy over its 6-arc min length.

Minimizing the $f/\#$ of the LUV/FUV Rowland-like spectrograph was a major trade during its design. As discussed in Sec. 3.2, an $f/10$ or faster beam in the Rowland circle was desired to ease the OTA requirements on the image formed by the OTA. During the design process, it was found that the image quality correction was unattainable from the aberration-corrected gratings at $f/10$ or faster. In fact, we found that $f/17.5$ (or slower) was required to achieve the $R \sim 20{,}000$ image quality. Optimization of the relay mirror aided this solution; hence, a sharp entrance slit image on the Rowland circle was not required. This design allowed the NUV and LUV/FUV to share a common entrance slit array at the TMA focus.

One of the two LUV/FUV gratings is selected by a mode select mechanism which inserts and mechanically registers the precise optical alignment of the selected grating. The full spectral region of 100 to 180 nm is split into two portions, one per grating. The current baseline design splits the spectra into two bands, one from 115 to 150 nm and the other from 140 to 180 nm, respectively. The baseline uses low-risk HST Lyman-$\alpha$-optimized Al + $MgF_2$ reflective coatings on the TMA mirrors and LUV/FUV relay mirror and gratings. A possible future option, dependent on the successful development of environmentally stable LUV Al + LiF coatings for these mirrors and the LUV grating, splits the spectra into two bands, one from 100 to 142 nm and the other from 130 to 180 nm, respectively.[13,14]

The LUV/FUV mode of the spectrograph uses an open CsI solar blind, curved MCP detector with 22-$\mu$m effective resels width. The active region of the MCP is 200-mm long by 70-mm wide. The MCP is housed in a vacuum enclosure with a door that is opened on-orbit. The detector operates at room temperature. The effective resel width of the 22-$\mu$m FPA resels imaged at the TMA focal surface by the LUV/FUV optics is 173 mas in-field.

The PSS modes share one entrance slit which is located in the TFA. This long L1 slit is 1.0 arc sec wide ($Y$) by 360 arc sec long ($X$).

## 2.6 *Wavefront Sensing and OTA Collimation*

Figure 8 shows the combination of OTA plus WFSs. Wavefront sensing will be performed occasionally to assess the optical alignment quality and image fidelity of the OTA. If alignment tune-up is indicated, the SM alignment will be adjusted deterministically to correct the OTA optical alignment and, therefore, the image quality of the OTA will be restored. When wavefront sensing is performed, light to the SIs is blocked by a PIM that rotates into the region between the OTA TM and the TMA focus near the OTA exit pupil. This mirror reflects the OTA field to the five WFS detectors. They determine the size and nature of misalignment-induced wavefront error (WFE) in Zernike coefficients, thus providing error signals to adjust the SM deterministically in the five degrees of freedom necessary to optically realign the OTA.

Adjustment of the OTA SM allows correction of limited image errors—third-order coma, defocus, and image plane centering, and tilt. To correct the whole OTA FOV, WFE must be sensed at multiple field points. For example, coma sensed at one field location, say due to one SM alignment error, such as decenter, could be completely compensated by tilting the SM, while increasing aberrations over the rest of the field. Our design study determined that five WFSs distributed over the full reach of the OTA FOV is the minimum number required to break this degeneracy.

Each WFS includes a lens which images the OTA entrance pupil onto its own WFS detector. Each WFS contains a detector array, with each detector sampling a portion of the full pupil, whereas the full array samples the entire pupil. For example, if each WFS is a Shack–Hartmann type that samples the pupil with a $50 \times 50$ array of sensors, the wavefront may be characterized at up to 25 cycles per aperture, assuming Nyquist sampling. This is more than sufficient sampling to sense third-order coma, focus, and image plane tilt, and decenter. The WFS technology has not been selected as of this writing. It could be a Shack–Hartmann, or perhaps a shearing interferometer, or one based on some other technical approach.

## 2.7 *Wavelength Calibration and Flat Fielding*

Figure 9 shows the in-flight WCS together with the FGS channels. Each FGS is adjacent to the stable member and uses a pick-off mirror near the OTA Cassegrain focus. The PSS takes spectra of astrophysical sources at large spectral resolving power over a wide wavelength range. Identification of the specific wavelength in the spectrum is required for identification of species, giving rise to specific spectral lines. Thus, CETUS includes a system to provide precisely known wavelengths.

When wavelength calibration of the PSS is performed, light to the SIs is blocked by a CIM that rotates into the region near the Cassegrain focus. This mirror reflects light from the wavelength calibration source to and through the PSS entrance slit. The spectrograph optics then disperses and images the known calibration spectrum. The design uses a STIS-HST-type Pt/Cr-Ne hollow core lamp. Again, FGS fine-pointing control is maintained during this operation.

In addition, light from flat-field sources flood each FPA to determine sensitivity and gain variations and stability of the detector pixels/resels. Flat-field lamps are mounted within each SI, respectively, to directly illuminate the relevant detector. A Xe line lamp, similar to the one used on GHRS-HST, will be used for the FUV channels, whereas Kr or deuterium line lamps like those used on STIS-HST are planned for the NUV channels.

## 2.8 *Fine Guidance Sensor*

Again, Fig. 9 shows the layout of the in-flight WCS together with the FGS channels. Figure 10 illustrates the dual FGS channels, their interface with the OTA interface, and FGS optical layout.

Two FGSs, each with a $1127 \times 1127$ arc sec FOV, provide fine-pointing error signals to the spacecraft ACS which in turn body points the OTA line-of-sight pointing jitter to <40 mas ($1\sigma$) jitter.

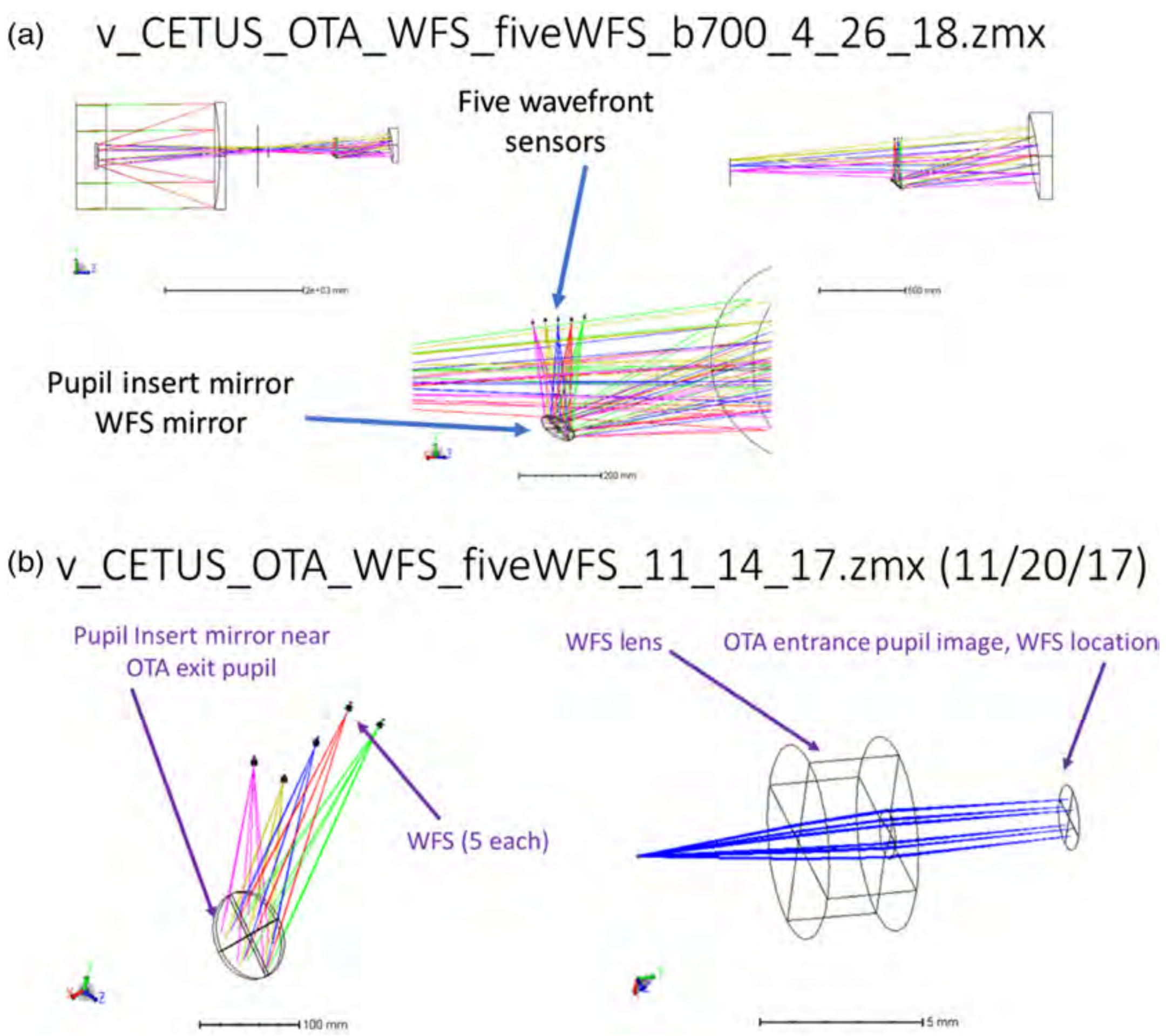

**Fig. 8** (a) OTA+ WFS, emphasizing PIM and WFSs. (b) OTA + WFS, additional view of locations of five WFSs.

As illustrated in Fig. 1, the two FGSs view the OTA FOVs separate from those of the SIs, thus enabling LOS pointing stability during scientific observations.

Each FGS uses an H4RG CMOS $4096 \times 4096$ pixel FPA with $10 \times 10$ $\mu$m pixels at $f/5$ with angular pixel width of 275 mas/pixel. Each FGS field, after folding by its fixed plano pick-off mirror near the Cassegrain focus, is imaged by lenses onto its FPA.

### 2.9 *Coatings*

Each mirror, grating, and window will use high-efficiency coatings over their respective wavelength ranges. Each fused silica window and order sorter will be coated with a single layer antireflective film. Lyman-$\alpha$-enhanced $Al + MgF_2$ reflective coatings, as used on HST, will be used on the mirrors of the CAM, the MOS, and the NUV PSS, and on the gratings of the MOS and NUV PSS. The dichroic beamsplitter in the MOS will reflect the NUV spectrum while transmitting visible light to the light trap to reduce the red leak in the spectrograph spectrum. The OTA mirrors and the LUV/FUV PSS select mirror and gratings will be coated with either the HST-type $Al + MgF_2$ or the LUV-enhanced $Al + LiF$, if the technology for a stable coating of the latter is demonstrated, thus enabling the LUV option for spectroscopy down to 100 nm.[13,14]

## 3 Optical Performance

### 3.1 *Design Image Quality and Spectral Resolving Power*

The designs presented here were developed using a combination of tools. The first-order parameters for the OTA and spectrometers were first modeled in Excel® spreadsheets using closed-form algebraic equations—a common approach to optical design. The first-order parameters of the OTA[7] were readily determined in this way. The spectrometer parameters were traded-off with closed-form expressions for spectral dispersion and spectral resolving power. These models calculate all first order parameters that define the spectrographs including angle of the incidence and diffraction orders of the echelle, spectrum length of each order, and order separation induced by the cross disperser. The model imposes the Wadsworth condition onto the cross disperser (i.e., choose angle of incidence so the midband wavelength is diffracted near zero angle of diffraction). Similar logic was applied to the CAM and MOS. In all cases, the resultant parameters were modeled in Zemax® for optimization and verification of the Excel® closed-form models. Each SI and each support system (WCS, WFS, and FGS1 and FGS2) were modeled and optimized in Zemax®. The Zemax® models were exercised to predict image quality for the nominal system and to develop optical tolerancing coefficients for performance error budgets (these are described in the next section)

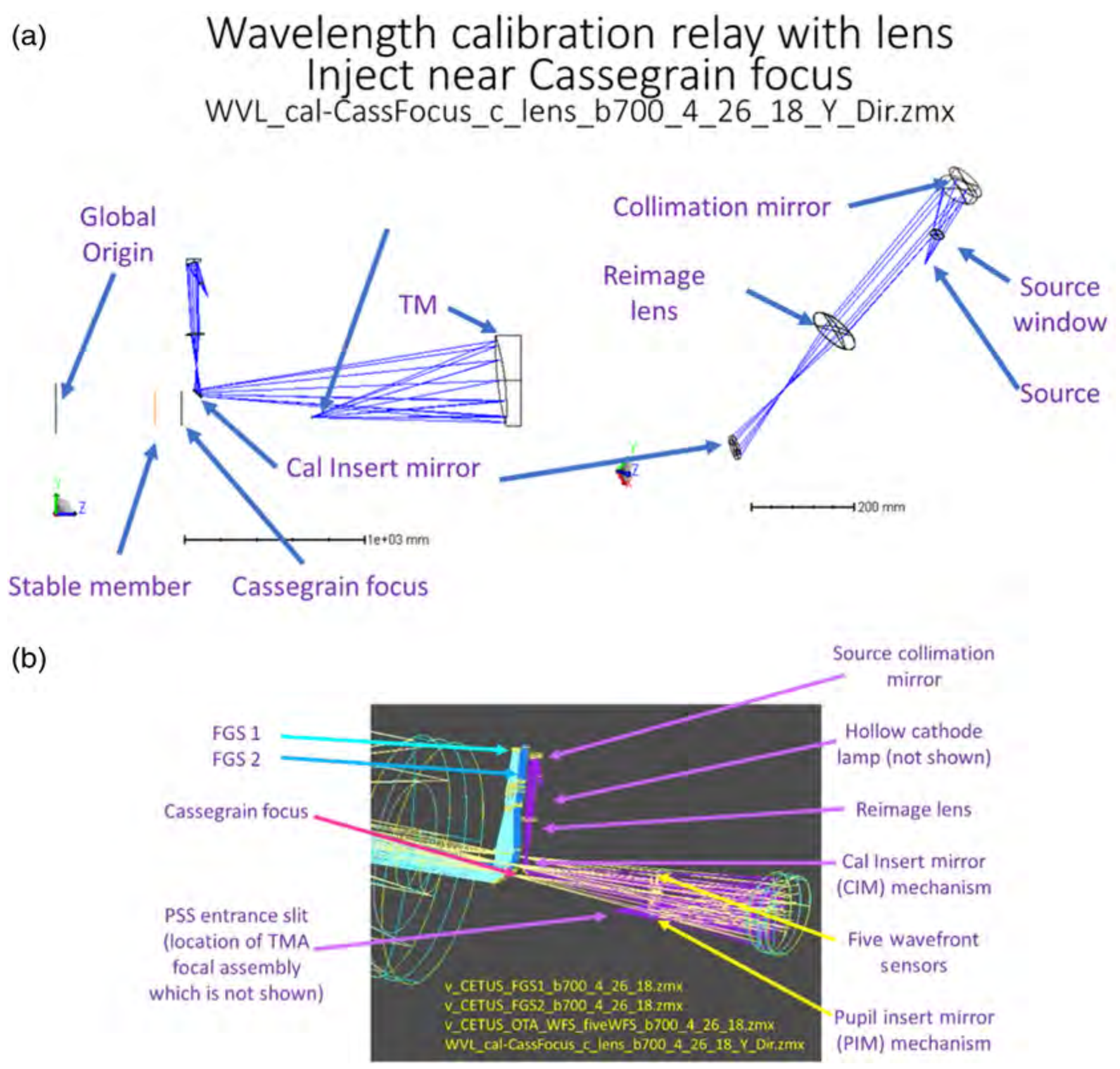

**Fig. 9** (a) In-flight WCS. (b) In-flight WCS with two FGS channels.

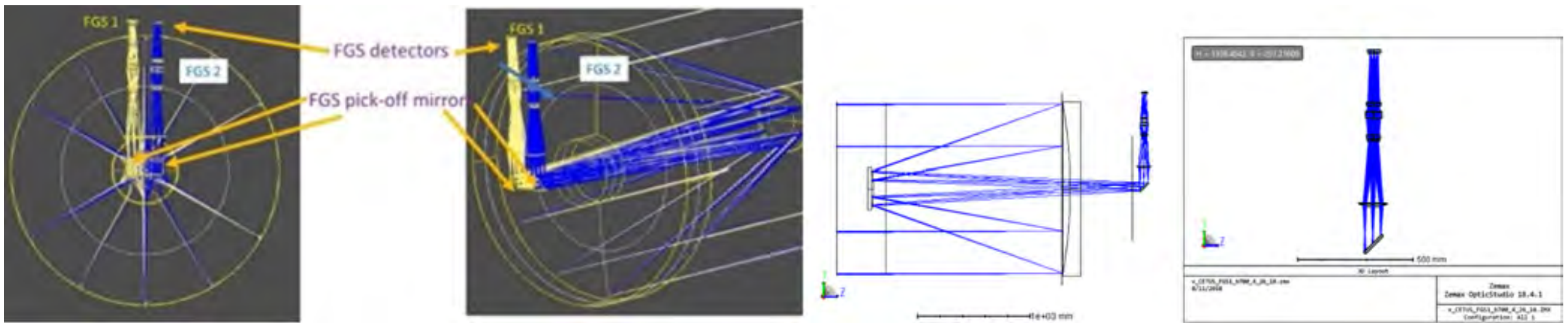

**Fig. 10** FGS channels and OTA interface: FGS 1 optical layout (right).

and to develop requirements for mechanisms. The Zemax® models generated STP files for computer-aided design mechanical mounting and packaging.

Summaries of the predictions of Zemax® and Excel® models of imagery, spectral dispersion, spectral resolving power, and the echelle-mode echellogram footprint are presented in the sections which follow.

#### 3.1.1 *Multiobject spectrograph*

Figure 11, spot diagrams generated in Zemax®, shows that the MOS images are well corrected and will achieve 1-pixel spectral resolving power of $R \sim 1000$ over the full FOV and the full spectral range. Figure 11(b) plots wavelength versus 1-pixel spectral resolving power demonstrating $R \sim 1000$ over the spectral band. Figure 11(c) shows that the full spectra for every object within the complete field is imaged within the $4\ \mathrm{K} \times 4\ \mathrm{K}$ 12-$\mu$m pixel array. Note that the image quality, image layout, and resolving power satisfy all requirements.

Figure 11(a) presents the geometric optics ray trace plots of two wavelengths ($\lambda$ and $\lambda + \Delta\lambda$) corresponding to $R = (\lambda/\Delta\lambda) = 1000$, thereby showing that the dispersion and image quality are adequate to achieve a 1-pixel resolving power over the whole FOV and over the entire wavelength band of the MOS. This performance metric is somewhat compromised at the short wavelength end of the spectrum between 180 and ~200 nm.

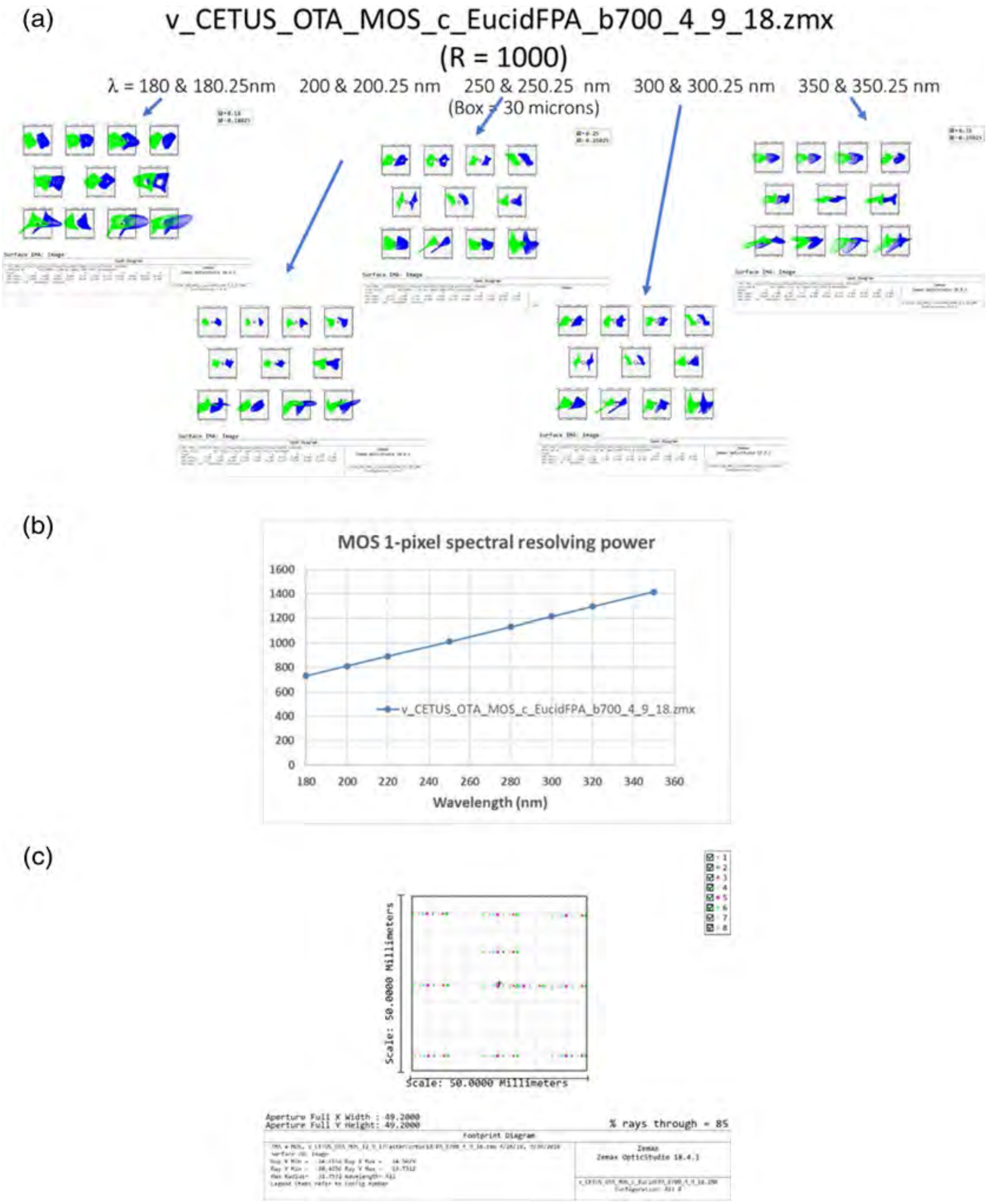

**Fig. 11** (a) MOS image quality over full FOV and spectral range for wavelengths separated by $R = 1000$, (b) 1-pixel spectral resolving power, and (c) full field and wavelength footprint on detector surface.

#### 3.1.2 *Camera*

Figure 12(a), geometric optics ray trace spot diagrams and physical optics diffraction encircled energy plots generated in Zemax® for the FUV and NUV CAM, demonstrates that the images are well corrected over the full FOV for monochromatic light. Figure 12(b) shows that the lateral color across the spectral filter bandwidths for the two wavebands somewhat degrades the imagery.

#### 3.1.3 *Point/slit spectrograph*

The geometric optics spot diagrams in Fig. 13(a) demonstrate that the NUV PSS image quality is well corrected over the full spectral range and that the NUV PSS achieves a 2.5-pixel $R \sim 40{,}000$ spectral resolving power. Figure 13(b) presents the image geometry of the echellogram displaying all diffraction orders. Order separation enables interorder sampling by >51 pixels. Figure 13(c) shows that the dispersion and image quality of both LUV/FUV G120 and G150 modes achieve $R > 20{,}000$ and imaging spectroscopy over a 6-arc min length field.

### 3.2 *Image Quality Error Budget*

#### 3.2.1 *OTA tolerances derived from SI resolution requirements*

The 1.5-m diameter CETUS OTA images a wide stellar field at the TFA at f/5. The OTA forms a sharp aberration-corrected

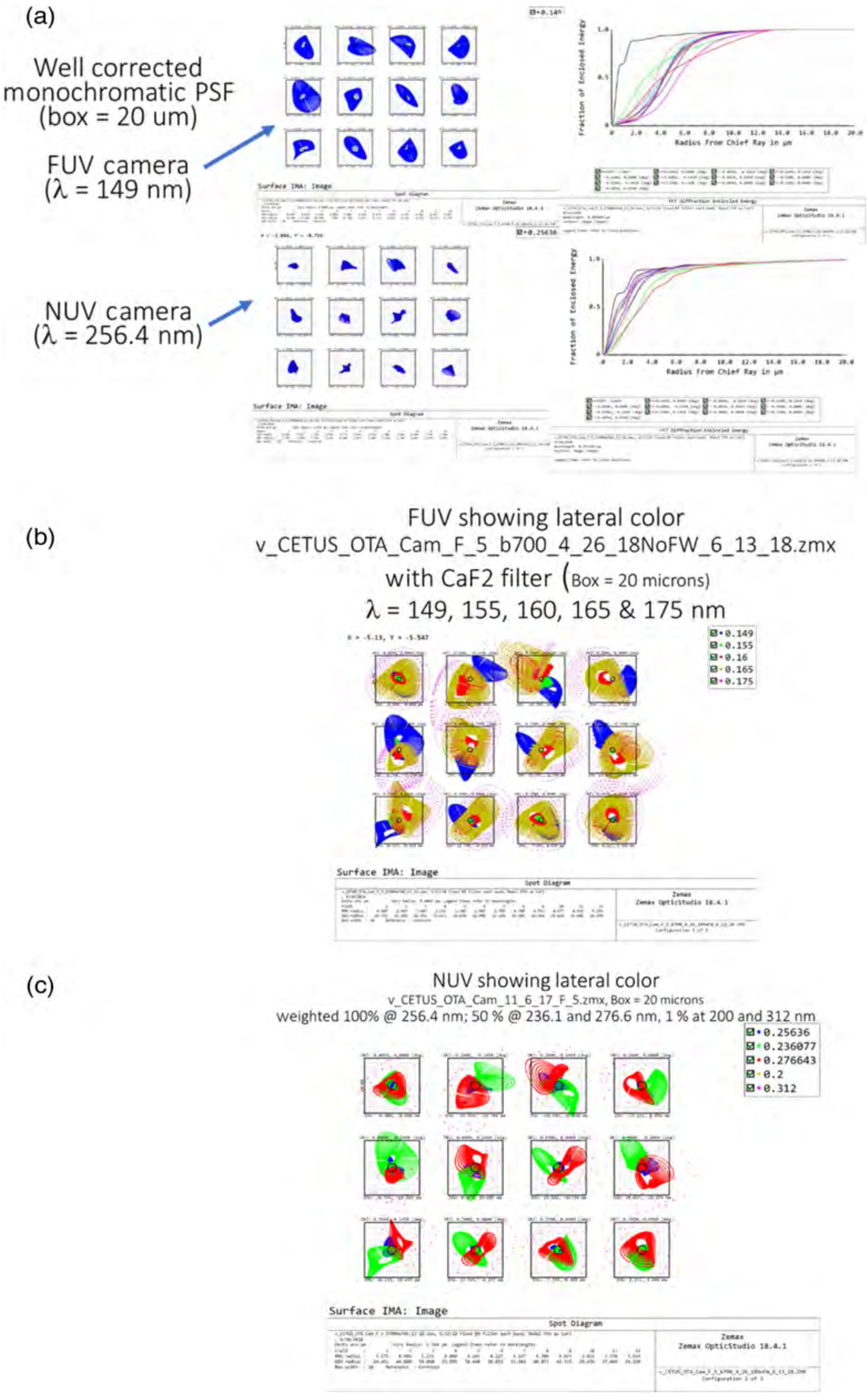

**Fig. 12** (a) FUV and NUV CAM image quality is well corrected over full FOV for monochromatic light. (b) FUV (top) and NUV (bottom) CAM image quality over full FOV in polychromatic light over spectral width of fixed bandpass filters exhibits residual chromatic aberration.

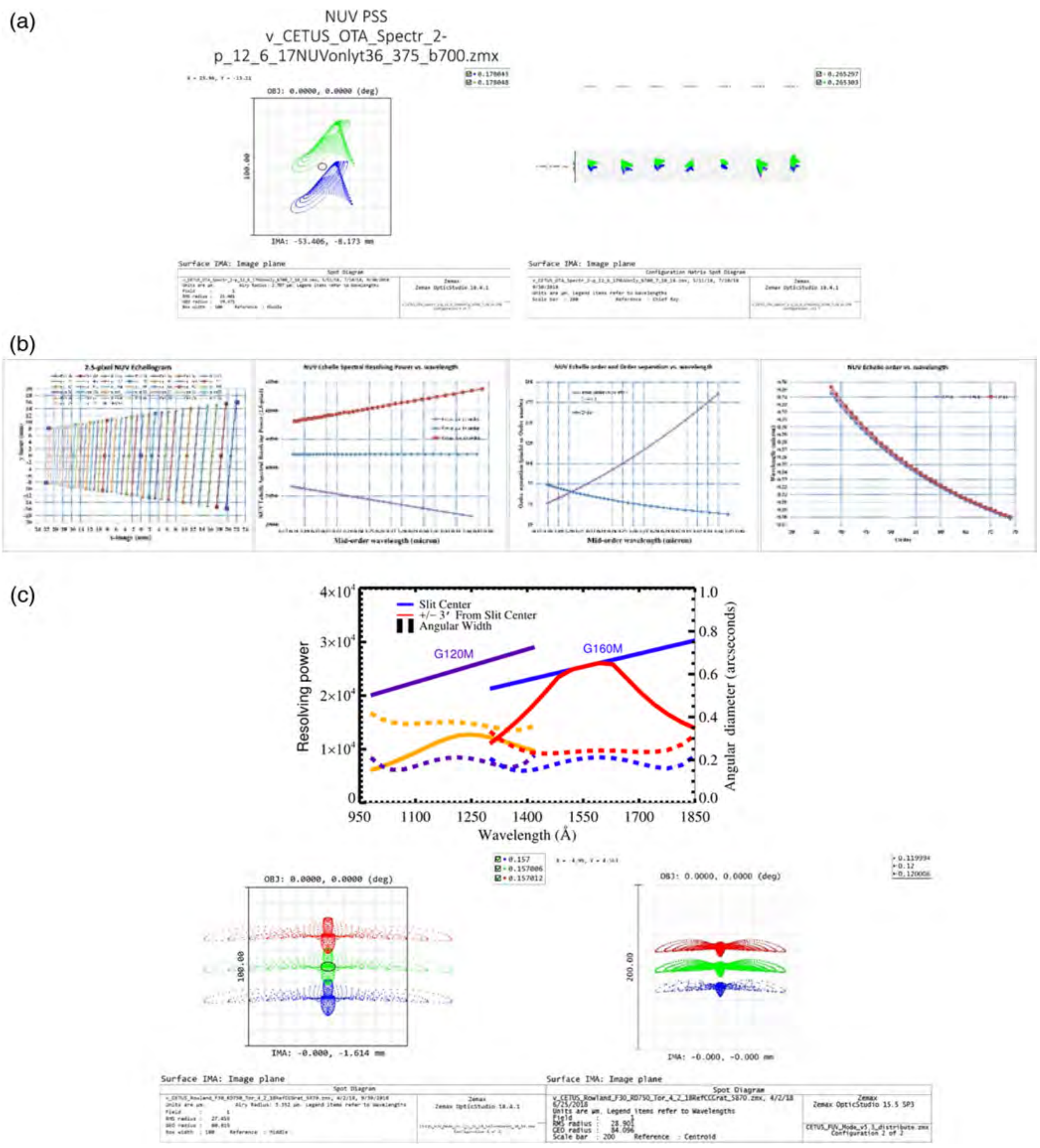

**Fig. 13** (a) NUV PSS spot diagrams show $R \sim 40{,}000$ spectral resolving power. (b) NUV PSS NUV left to right: echellogram, spectral resolving power versus wavelength, echelle order and order separation versus wavelength, and echelle order versus wavelength (bottom) to allow sampling of interorder background. (c) NUV PSS LUV/FUV spectral resolving power $R > 20{,}000$ versus wavelength (top). Despite large cross-dispersion flaring, most of the energy is contained within ~2 MC Presels (44 micron) (bottom). [Design by B. Fleming, LASP, University of Colorado].

image of the science field at the TMA focus with effective focal length of 7.5-m with the conic constants of three mirrors selected to correct to third-order spherical aberration, coma, and astigmatism. The OTA image surface is plano with curvature corrected to zero Petzval curvature by choice of mirror radii of curvature and axial spacing. In addition, each of the three SIs reimage the OTA image to their respective detector.

As part of the design process, image quality error budgets were developed combining errors from individual components and allocating tolerances to meet the overall system performance metrics. The budgets were developed in two interrelated categories: image quality delivered by the OTA at the TMA focus and image quality delivered to the detector surface of each instrument by the instrument relay. To derive the requirements for OTA image quality over the TMA focus surface, the spatial resolution requirements for each SI at the OTA focal surface was derived. The effective spatial resolution width (ESRW) at the TMA focus, $w$, is the image width of the FPA resolution sampling width, $s$, as imaged by the SI optics image to the TMA focus, respectively. If $F_{\text{OTA}}$ is the $f/\#$ of OTA and $F_{\text{SI}}$ is the $f/\#$ of a SI at its FPA, we find:

- For the CAM: $w = s \cdot F_{\text{OTA}}/F_{\text{SI}}$
- For the MOS and PSS: $w = s \cdot (F_{\text{OTA}}/F_{\text{SI}}) \cdot (\cos\alpha/\cos\beta)$, where $\alpha$ = the angle of incident at the grating and $\beta$ = the angle of diffraction accounting for the anamorphic magnification of the gratings in the spectrographs.

We define a parameter, $m_{\text{SI}}$, the effective magnification of an instrument, to be the ratio of the image size at the detector of the instrument to the image size at the TMA focus. This factor is therefore $m_{\text{SI}} = s/w$, which includes anamorphic magnification by the diffraction gratings in the spectrographs.

The $f/5$ FUV and NUV CAM modes use an effective resolution width at the detector of 1 pixel (NUV) or 1 resel (FUV), so that $s = p = 20$ $\mu$m resels for the FUV CAM and $s = p = 12$ $\mu$m for the NUV CAM, where $p$ is the physical width of a pixel or a resel. The MOS and LUV PSS also use 1-resel sampling, so $s = p = 12$ $\mu$m for the MOS and $s = p = 22$ $\mu$m for the LUV PSS.

Note that in Sec. 3.2.4 of this paper we mention that the LUV PSS stresses the requirements on the OTA image quality. To ease the requirements on the OTA image quality, the NUV PSS design increased the dispersion and adopted $s = 2.5$ pixels $= 30$ $\mu$m, instead of using 1-pixel resolution width. The large format CCD provides sufficient format size to capture the resultant echellogram. Figure 14 summarizes the resultant ESRWs at the TMA focus for each instrument and the "effective diffraction-limited wavelength" if the value for $w$ is the FWHM of an effective Airy pattern (see Sec. 3.2.2).

### 3.2.2 *OTA image quality metrics*

We consider two distinctly different, but relatable, models of the OTA point-spread function (PSF) for the image quality error budget. Using these we can quantify the errors due to a variety of sources and combine them to assess how various errors contribute to the overall system image quality and to predict the properties of resultant image PSF. The first approach employs a geometric optics approximation to light propagation and image formation. Applying geometric optics, image errors are modeled as Gaussian functions with FWHM, $\delta$, and standard deviation, $\sigma$. The FWHM at the TMA focus is equated with the parameter, $w$, defined above, so $w = \delta$. Assuming that the individual errors are statistically independent, the resultant FWHM values are combined statistically in a root-sum-squared fashion. The second approach is a physical optics approach where it is assumed that, even though the system is not diffraction-limited at its operational wavelengths, it will be diffraction-limited at some "effective wavelength." The "effective wavelength" at which the system is diffraction-limited is derived together with the FWHM of the diffraction Airy PSF at that wavelength. Then, that value is interpreted to $w = \delta$. Following Maréchal[15] the resultant wavefront errors are combined statistically in a root-sum-squared fashion. We combine the resultant FWHM statistically with the values for other errors determined by the geometric optics model.

We only apply this physical optics technique for specification of the OTA mirrors, PM, SM, and TM, including wavefront errors resulting from fabrication, mounting, and from the reflective coating with the mirror at room temperature in a laboratory setting.

*Geometrics optics approach using Gaussian PSF.* Using a Gaussian model, the two-dimensional PSF is modeled as a function of spot radius, $r$, as $f(r) = e^{-0.5z^2}$, where $z = \frac{r}{\sigma}$. With $r_{\text{HM}} = 0.5\,\delta$, defined by $f\,(r_{HM}) = 0.5$, we find $\sigma = \frac{\delta}{2\sqrt{2\,\ln(2)}}$. Therefore, the FWHM $= \delta = 2\sqrt{2\,\ln(2)}\,\sigma = 2.355\sigma$.

*Physical optics approach: "Effective wavelength" for "Diffraction limit."* In the physical optics model of contributors to the PSF size, we first allocate values of WFE to each mirror of interest. We then combine these mirror figure errors by root-sum-square of the individual mirror errors. We convert the resulting WFE into FWHM by first determining the pseudowavelength, $\lambda_{\text{DL}}$, at which this WFE would yield a "diffraction-limited" PSF. The FWHM of this PSF becomes the effective FWHM that is then combined with other FWHM values by root-sum-square to estimate the system FWHM.

In the wavefront error physical optics image model, the image profile is an Airy function with a FWHM of $\sim t(\frac{\lambda_{\text{DL}}}{D})$,

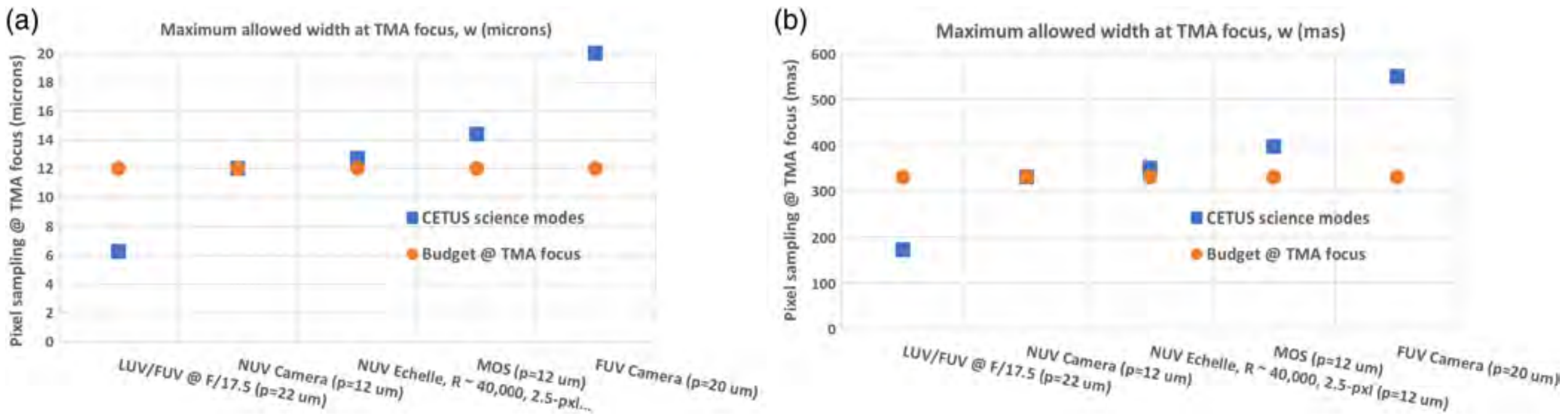

**Fig. 14** (a) ESRWs at the TMA focus for each instrument and (b) the "effective diffraction-limited wavelength" if the value for $w$ is the FWHM of an effective Airy pattern.

**Table 2** The resulting OTA mirror fabrication and mounting error budget with specific definition of spatial frequency limits for each spatial frequency region.

| | Mirror overall (WFE nm rms) | 63.3 | 32.0 | 32.0 |
|---|---|---|---|---|
| | | PM | SM | TM |
| | Component tolerances (mounted) | nm rms, surface | | |
| Fabrication and mounting | LSF: <5 cpa | 31.3 | 15.3 | 15.3 |
| | MSF: 5 to 20 cpa | 3.0 | 3.0 | 3.0 |
| | HSF: >20 cpa | 1.5 | 1.5 | 1.5 |
| | Microroughness: <1 mm | 1 | 1 | 1 |
| | Coating (nm rms surface) | 3.0 | 3.0 | 3.0 |
| | Overall coated (surface) | 31.6 | 16.0 | 16.0 |

where $\lambda_{\mathrm{DL}}$ is an effective pseudowavelength at which the OTA will be "diffraction-limited," and $t$ is a parameter. Ignoring the diffraction by the spiders but including a centrally obscured aperture with linear obscuration ratio, $\varepsilon$, the "diffraction-limited" PSF as a function of image radius, $r$, is $\mathrm{PSF}(r) = \left[\frac{2}{(1-\epsilon^2)}\right]^2 \left[\frac{J_1(b)}{b} - \epsilon^2 \frac{J_1(\epsilon b)}{\epsilon b}\right]^2$, where $b = \frac{\pi r}{8}$, and $J_1(b)$ is a first-order Bessel function. With the central obscuration of the OTA, the parameter $t = 0.972$ and the $\mathrm{FWHM} = \delta = 0.972 \frac{\lambda_{\mathrm{DL}}}{D}$. Defining "diffraction-limited" as allowing a WFE of ~0.0752 waves RMS (Strehl = 0.8) at the effective wavelength, the allowed wavefront error $\omega = 0.0752\ \lambda_{\mathrm{DL}}$ becomes $\omega \sim 0.077 D\,\delta$.

To verify this physical optics model, the use of a "pseudo-diffraction-limited wavelength" must be justified. To this end, a geometric optics surface error model has been developed for the PM and the predictions of FWHM by the two models are compared. If in the physical optics model, 63.3-nm rms WFE is allocated to the PM, i.e., ~0.1 wave RMS at 632.8 nm, the PM would be diffraction-limited at $\lambda_{\mathrm{DL}} = \mathrm{WFE}/0.0752$. The optic PSF would be an Airy pattern at that wavelength. Then the FWHM at the TMA focus would be $0.972 * (\mathrm{WFE}/0.0752) * (f/D)$, 4.09 $\mu$m for this example, where $f$ and $D$ were previously defined. The FWHM for the geometric optics model is $4 \times 2.355\ \alpha f(D_p/D)$, where $\alpha$ is the RMS slope error and $D_p$ is the diameter filled by an on-axis field source. Evaluating we find that the geometric FWHM is 4.09 $\mu$m, if $\alpha = 1.16 \times 10^{-7}$ radians RMS. If this slope error continued uniformly across the full diameter, the error in height would be $\alpha * D_p \sim 174$ nm, which gives a FWHM $\delta$ of ~74 nm rms applying the factor 2.355, as discussed above. This is within 16% of the assumed 63.3-nm RMS WFE and confirms the model to this accuracy.

### 3.2.3 *OTA image quality error budget*

*Contributors.* To evaluate how errors in the OTA wavefront contribute to each SIs image quality, error budgets are also constructed for each SI. Fabrication, alignment, and operational error sources are considered, including those due to OTA design residual, OTA mirror fabrication and mounting, OTA mirror alignment, line-of-sight jitter and roll jitter, environment effects such as temperature-induced mirror misalignment and mirror distortion, and coating thickness variation. As the design process is in the early phases with still many unanswered questions, which need extensive analyses beyond the current scope of this study, two additional categories are included: "reserve" and "other."

In the current effort, particular attention is paid to the two categories to which the system is most sensitive: OTA mirror fabrication and SM alignment. Mirror fabrication errors are allocated based on an assessment of the current state-of-the-art mirror fabrication tolerances for a cost-effective program that will still achieve superior UV performance. For this category, the wavefront error is parsed into distinct spatial frequency regimes referenced to the beam footprint size of a single field point on each mirror: low spatial frequency (LSF) <5 cycles per aperture (cpa), midspatial frequency (MSF) 5 to 20 cpa, high spatial frequency (HSF) >20 cpa, and microroughness with spatial periods <1 mm. The resulting OTA mirror fabrication and mounting error budget are presented in Table 2.

Note that a looser tolerance is budgeted for the PM as it is the most difficult mirror to fabricate due to its size. Also, the MSF is budgeted to be four times looser than that achieved by the HST. These design considerations help in reducing the fabrication costs for the OTA.

Table 3 summarizes the general categories of errors within the current budget with their present allocations as previously defined. This table highlights the sensitivity of the image error to misalignment of the SM and particularly to the SM despace (i.e., axial) error. It completely dominates this error. Despace contributes ~107-nm rms WFE, ~88% of the total SM misalignment (including tip, tilt, and decenter) error of 121 nm rms WFE. When combined statistically by root-sum-square, the nondespace alignment error term is 57.7-nm rms.

Note that the assumed alignment errors are not the misalignments from before launch to operation. Instead they are the allowed alignment drift between each periodic on-orbit realignment of the SM and the next SM realignment. The optical, mechanical, and thermal design of CETUS will stress the maintenance of long-term stability to greatly reduce the need for realignment and to extend the time between realignment.

SM despace, defined as the error in axial location of the SM, defocuses the image formed at the TMA focus. The TMA image size is very sensitive to despace due to the large magnification of the PM focal length by the combined SM and TM. Despace limits the allowed drift in alignment after realignment. The 5-$\mu$m despace allocation is feasible but will drive OTA stability trades. The next largest error is WFE induced by the PM fabrication/mounting errors. Jitter contributes lesser amounts. The general categories on “environment,” “other,” and “reserve” carry fairly large allocations in the current budget. The budgeted errors included in this allocation are shown.

**Table 3** General error categories within the current budget with their present allocations and key component allocations particularly SM misalignment tolerances after in-flight realignment by SM adjustment.

| Tolerance allocation | WFE (nm rms) | rms spot dia. microrad |
|---|---|---|
| OTA design residual | 34.8 | 0.30 |
| Fabricate PM mounted | 63.3 | 0.23 |
| Fabricate SM mounted | 32.0 | 0.12 |
| Fabricate TM mounted | 32.0 | 0.12 |
| Due to SM alignment errors[a] | 121.2 | 1.04 |
| LOS jitter with roll jitter | 22.5 | 0.19 |
| SM tilt jitter | 11.3 | 0.10 |
| Environment | 50.0 | 0.43 |
| Other | 50.0 | 0.43 |
| Reserved | 50.0 | 0.43 |
| [a]Due to SM despace | 106.6 | |
| **Allocated SM alignment and LOS jitter error** | | |
| Despace | 5 | micron |
| SM tilt | 5 | arc sec |
| SM d*x*, d*y* | 20 | micron |
| SM tilt jitter | 50 | mas |
| LOS jitter/axis | 40 | mas |
| LOS jitter with roll jitter | 40 | mas |

The sensitivity to despace highlighted in Table 3 can be readily understood. Owing to the longitudinal magnification, defocus at the TMA focus $df$ becomes $df \sim d_2 * (m_2^2 + 1) * m_3^2$, where the magnification of the SM $m_2 = 5.74$ and of the TM $m_3 = -0.60$ and $d_2$ is the SM axial error, i.e., despace. Note that the error varies as the square of the mirror magnifications causing high sensitivity to mirror axial locations. Evaluating df $\sim d_2 * 12.3$. Therefore, a despace error of 5 $\mu$m introduces defocus at the TMA focus of 61.5 $\mu$m. At $F = 5$, the $f/\#$ of the OTA, the resultant geometric image diameter is ~12.3(df$/F$) in microns. Thus, to this approximation, 5-$\mu$m SM despace introduces a geometric image diameter of 12.3 $\mu$m, a value larger than the whole system budget and therefore drives the design. Fortunately, the actual effect as determined by error analysis using Zemax® models is ~6.9 $\mu$m for a 5-$\mu$m despace, about ½ of the simple first-order estimate of 12.3 $\mu$m. This is because Zemax® includes all system ray trace effects. The presented error budgets use the Zemax® results.

#### 3.2.4 *OTA image error budgets*

Table 4 summarizes the estimated FWHM for each SI at the TMA focus together with the derived requirement of each. The requirement for the NUV CAM is adopted as the system requirement. Note that all but the LUV/FUV $R \sim 20{,}000$ channels meet this requirement. Owing to its slow f/# at its detector, it must have a FWHM image width at the TMA focus of less than 173 mas to achieve $R \sim 20{,}000$. Scaling to the system budget allocated to the CAM, the LUV/FUV will be limited to ~10,000 for slit-less observing, if the image FWHMs were 330 mas. (As an option, a 173-mas wide portion of the 6-arc min long slit could truncate the extended PSF, losing energy, but allowing $R \sim 20{,}000$ long-slit imaging spectroscopy. This option represents an ongoing design trade as of this writing.) With these considerations in mind, we propose a single long slit, 6 arc min in length and 1 arc sec wide, possibly with a small region necked down to 173 mas for use on isolated stellar objects in the $R \sim 20{,}000$ LUV/FUV mode of the PSS.

#### 3.2.5 *System image error budgets: OTA + SIs*

The OTA error budgets in the previous section, which predict the quality of the image required by each SI, are combined with errors due to imperfections of each optical element in each SI, respectively, to estimate the FWHM of each image at the SI detectors. This budgeting process uses the geometric optics approach described above. Also included are the SI errors in categories of alignment, other, and reserve, providing error estimates that will be allocated to component design in later design trade-offs. The FWHM estimates from each instrument are combined with the above OTA error budget prediction, properly scaled by the effective magnification of that SI, using a root-sum-square algorithm.

Table 4 also summarizes our prediction that the FWHM of each SI combined with the OTA, except the LUV/FUV when it is operated in a slit-less mode where the 330 mas PSF from the OTA defines the effective entrance slit of the spectrograph, will meet the image quality requirement for their respective science goals.

One important insight from the OTA+SI error budgeting relates to the requirement on the RMS slope error of the mirrors in each SI. The CAM mirrors have a tighter specification on slope error (1.0 $\mu$rad RMS slope error) than required for the MOS and PSS mirrors (1.5 $\mu$rad RMS slope error). Based on the input from Arizona Optical Systems, these slope specifications are within the current state-of-the-art for mirror fabrication processes.

#### 3.2.6 *System and SI trades summary*

Key system trades, including the image error analysis, drive the optical system design. The MOS, CAM, and LUV/FUV spectrograph each use a single-pixel/resel, FWHM, resel. Dithering of the image on the focal plane is used to extract the required resolution. A single pixel as imaged at the TMA focus, therefore, sets the OTA tolerances. The NUV CAM, which uses CCD 12-$\mu$m pixel width, is used for the system baseline requirement. The $f/\#$ of the NUV PSS is relaxed to $f/15$ to achieve the aberration control needed for $R \sim 40{,}000$ spectroscopy. If the one-pixel criteria were applied to this channel, the demagnification of the pixel image at the TMA focus will require tighter OTA tolerances. To resolve the issue, a 2.5-pixel resolution criterion is adopted by increasing the dispersion. Fortunately, the CCD detector is physically large enough to accommodate the increased dispersion.

**Table 4** Budgeted and predicted image FWHM at the TMA focus and at SI FPA for each SI.

| | OTA error budget totals @ TMA focus | | | | | OTA + SI error budget totals, s_eff | | | |
|---|---|---|---|---|---|---|---|---|---|
| | Mas | | Micron | | | Micron | | | Micron |
| Mode | Budget | Predict | Budget | Predict | Complies? | Budget | Predict | Complies? | SI FWH |
| NUV CAM ($p = 12$ $\mu$m) | 330 | 282 | 12.0 | 10.2 | True | 12.0 | 11.9 | True | 6.11 |
| Design requirement | | | | | | | | | |
| FUV CAM ($p = 20$ $\mu$m) | 550 | 282 | 20.0 | 10.2 | True | 20.0 | 11.9 | True | 6.06 |
| MOS ($p = 12$ $\mu$m) | 396 | 282 | 14.4 | 10.2 | True | 12.0 | 11.0 | True | 6.87 |
| LUV/FUV, $f/17.5$ ($p = 22$ $\mu$m) | 173 | 282 | 6.3 | 10.2 | False | 22.0 | 36.5 | False | 6.62 |
| NUV echelle, $R \sim 40{,}000$, 2.5-pixel ($p = 12$ $\mu$m) | 349 | 282 | 12.7 | 10.2 | True | 30.0 | 25.3 | True | 7.40 |

The error budget shows the sensitivity of the image quality to SM misalignment, especially despace. Therefore, we include a mechanism to adjust the SM in-flight with a multisensor WFS to provide error signals to guide the adjustment. We have also implemented design measures to insure optical stability between realignments: low CTE materials, kinematic mounting of optics and instruments to minimize instrument distortions, and controls to minimize the effects due to changes in temperature levels and gradients.

## 4 Optical Alignment and System Test

The CETUS system design incorporates features driven by technical needs and programmatic concerns. Tools that enable full aperture, prelaunch, full system optical alignment verification are incorporated into the design. Recognizing that the OTA and each instrument may be designed and built by different entities, each instrument and the OTA are decoupled from each other. The OTA is one complete independent assembly that includes the PM, SM, TM, WFS, and TFA. Each SI, each FGS, and the wavefront calibration system are independent assemblies, respectively. These instruments will be assembled, aligned, tested, and calibrated before their integration into the OTA assembly.

To address programmatic concerns, the design ensures that, during integration and test, any one of these entities can be removed and reintegrated without affecting the alignment of the OTA or any other assembly. To that end, no portion of the OTA undergoes disassembly when installing or removing any other assembly. Therefore, if any component within any SI needs replacement or repair, e.g., a mechanism problem, a detector problem, or to clean a mirror, the remainder of the system assembly integration and test (AI&T) can proceed without any impact on the schedule.

### 4.1 *OTA Optical Alignment Test*

The design includes alignment fiducials referenced to the TFA to enable full aperture optical testing of the completed OTA. Figure 15 shows their physical location. These fiducials are the end termination of five single-mode fiber-optic strands, each providing a point source of light intimately referenced to the TFA. Each is conjugated to a WFS whose locations are shown in Figs. 1 and 15. During system optical alignment the fibers light output projects a full aperture beam through the OTA. The beam exits collimated from the front of the OTA to a full aperture flat test mirror that, in turn, reflects the beam back into the OTA which then images it at the TMA focus. The other end of each fiber terminates at the outer surface of the CETUS system at a location accessible during AI&T. This end of the fiber is accessed by the removal of a small cover. The single-mode fibers are 9-$\mu$m diameter fiber optics transmitting wavelengths longer than 320 nm. When used, the cover is removed and a suitable light source is placed to illuminate the fiber. The light then propagates through the fiber and then emanates from the fiber image end to fill the OTA aperture. As illustrated in Fig. 15, the FGS paths also have fibers for the same purpose. All fibers are integral to the design. Their external ends will be accessible even in the late stages of integration, perhaps even when the facility is mounted on the launch vehicle. We are yet to study the practicality of removing them before flight and the system issues that would arise if we fly them. The latter category, including contamination, stray light from light scattered from the fiber ends, and thermal conduction down their length, will be studied in later design phases.

### 4.2 *System Optical Alignment Test: OTA + SIs*

The non-OTA assemblies, MOS, CAM, PSS, FGS, WCS, and WFS, are separate modules that are aligned and tested individually and then integrated to the OTA assembly. Following the approach used in HST and other space systems, they are designed to meet the agreed-upon interfaces defined in ICD. An ICD is generated defining all interfaces among the OTA and the other instruments. The ICD defines each of the interfaces, including optical, mechanical, structural, mass, center of gravity, power, electrical, thermal, FPA, FPA cooling, ground, and in-flight calibration, environments, contamination, access to critical components, reliability, test requirements, alignment fiducials, and mounting tolerances.

Using dedicated OTA simulators, each SI is aligned with visible light using suitable surrogates for optics and detectors. Each assembly is aligned with an OTA simulator that is designed to properly simulate the field and pupil properties of the flight OTA as viewed by the respective SI. To ensure that the OTA simulators properly mimic the OTA, especially for the MOS and CAM, we also test these with the full aperture OTA autocollimation test using the fiber-optic sources as described above. This provides a double-blind verification of the OTA simulators and of the alignment of the total system, including pupil aberrations. We note that the OTA simulator for PSS can be a simple $f/5$ input using a small field paraboloid of revolution mirror because OTA pupil properties do not affect the PSS due to its very small FOV.

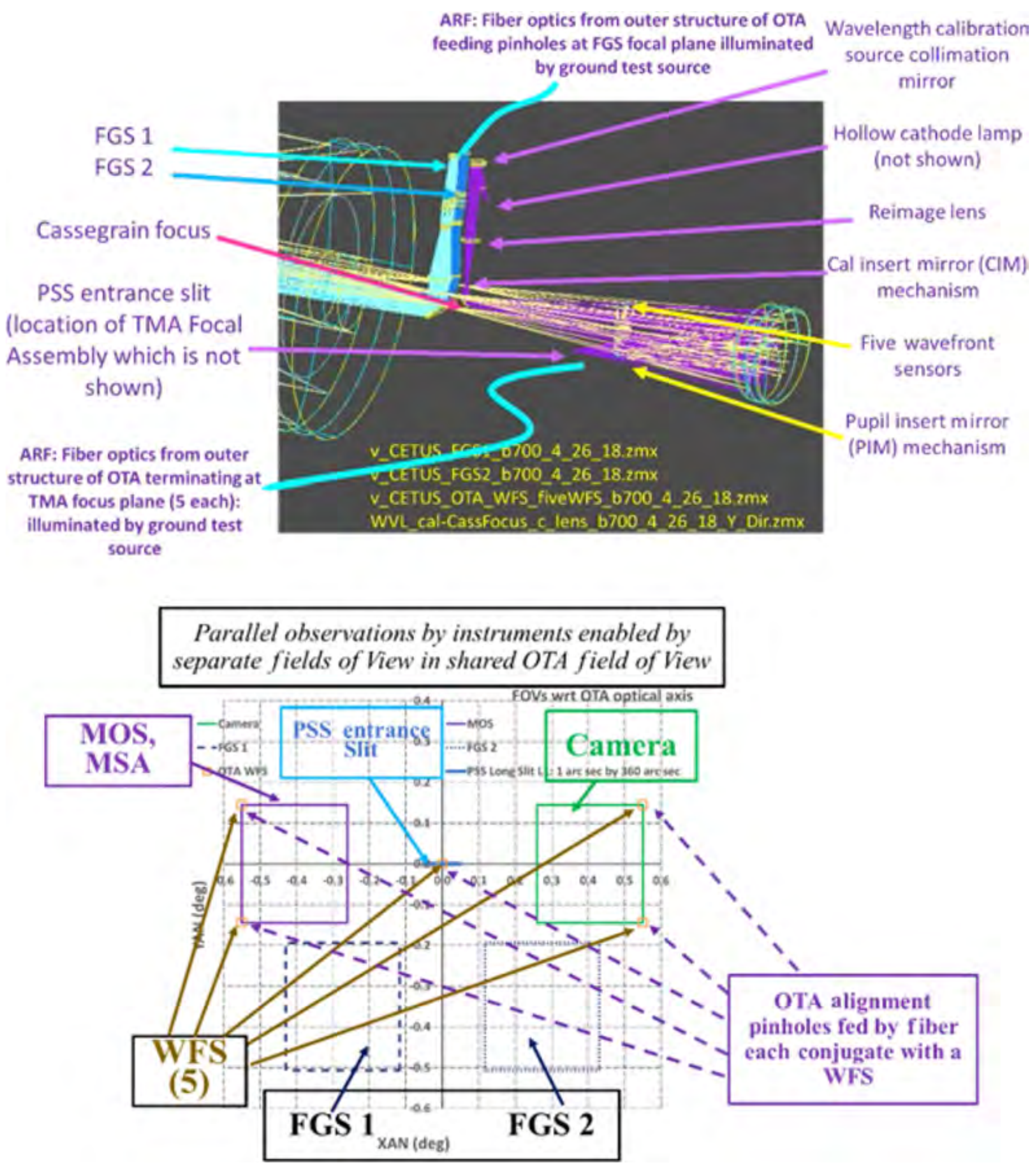

**Fig. 15** Illustration of location of alignment reference fibers.

## 5 Discussion and Conclusions

We have designed a probe-class mission concept called the CETUS. CETUS will continue space-based vacuum UV scientific exploration of the Universe beyond HST's lifetime. Its 1.5-m aperture wide FOV telescope with its complement of three SIs allows long exposure, efficient, wide-field surveys of stellar objects by both a NUV MOS spectrograph and a FUV/NUV CAM. The third SI has two high-resolving power spectroscopy modes: (1) a LUV/FUV/NUV, long slit, imaging spectrograph that extends the CETUS spectral capability to HST cut-off (115 nm) and potentially into the LUV (100 nm) and (2) a NUV point source spectrograph.

The comprehensive design includes all systems and features required to accomplish the full scientific observation mission. To that end, the design includes the OTA with WFSs, the above-referenced SIs, and all supporting sensors that the mission requires: FGSs, in-flight wavelength calibration sources for the PSS, and flat-field in-flight calibration sources for all SI detectors.

This paper describes the current optical design of CETUS and major trades leading to the design. Although we are still in an early design phase, we have completed sufficient design trades to yield a mature design. The system design that accomplishes the scientific objectives is a low risk with high heritage design. Our design trades and error analyses have balanced instrument performance to place reasonable tolerances on the OTA and all instruments. The highly efficient photon detection design is enabled by (1) using an all-reflective optical design of the OTA and of the instruments, (2) using Lyman-$\alpha$-optimized reflective mirror coatings (or as an option, LUV-optimized), (3) minimizing the number of reflections in each optical path, (4) using field-sharing that permits simultaneous observing by SIs, (5) selecting a continuous viewing orbit, and (6) using detectors with minimum number of windows and with high quantum efficiency. Of note, the LUV/FUV PSS design is specifically designed for maximum throughput by using a photon-efficient single-disperser design with heritage from HST's COS instrument.

To maximize science return, the design is specifically formulated to provide temporally efficient use of mission time. One key feature implements parallel observations by the MOS and CAM. Each instrument samples a dedicated separate portion of the large well-corrected OTA FOV. This approach to field-sharing enables simultaneous science observations by all instruments and thereby enables time-efficient surveys. Even during PSS observations of a selected object, serendipitous observations by the MOS and the CAM are allowed.

The selected Sun/Earth L2 orbit enables continuous, uninterrupted viewing of stellar objects. At any point in the orbit, a full hemisphere is potentially viewable. The full $4\pi$ steradian view of space is accessible over a 1-year time frame, making use of the Earth's orbital progression around the Sun.

High observational efficiency also results by maximizing spectral throughput and photon efficiency. The number of photon absorbing mirrors per channel are minimized by using no fold mirrors before the TMA focus, resulting in a relatively long optical system that is nonetheless completely compatible with the Falcon 9 launch vehicle fairing.

The separated field-sharing approach also allows parallel integration of the individual instrument modules. The resultant design specifically allows integration of each instrument without affecting the alignment of previously installed instruments or of the OTA. This approach reduces schedule risk during AI&T because an issue with any instrument can be addressed without affecting the alignment or integration status of any other system. It also provides programmatic options in that the instruments can be provided by external sources who will design to mutually agreed-on interfaces.

Design tolerances and their effects on system image quality are understood and are allocated across the system through application of systems engineering tools.

A FUV/LUV payload's functionality and longevity is based on contamination management. The design recognizes the necessity for minimizing the accumulation of contaminates during AI&T and flight on optical surfaces to retain efficiency. To this end, several design features. These include a full aperture door closed during orbit maintenance propellant firings and safing. For CETUS, we apply a four-method approach: (1) materials throughout will be selected to be low outgassing or to contain any small volume of volatile materials (e.g., epoxy); (2) all materials, especially the structure and bonds, will be baked to evolve nearly all volatiles, and maintained clean with dry $N_2$ or vacuum during storage or test. Steps 1 and 2 will minimize any contamination which could polymerize during service; (3) on commissioning in space, all optical surfaces will be radiatively warmed to redistribute any contamination, and periodically rewarmed to the redistribution temperature, and (4) all optical surfaces will be maintained in a warm biased state, $\sim +5$C warmer then neighboring areas, to ensure that the optic does not act as a cold finger and collect residual contamination. Otherwise, the optics would radiatively couple to cold space and pump contamination toward them. These are the classic measures of maintaining cleanliness of spaceborne UV instruments.

To ensure success after launch, the optical design is fully testable on the ground before launch with features built in to enable such testing. The CETUS design includes alignment sources such as the ends of single-mode fiber optics precision mounted on the TFA which are strung to ports residing on the external skin of the fully integrated flight article. When an external light source feeds the external port end of the fiber, the fiber end emits light that is then collimated by the OTA to exit the front end of the OTA. During test, an external full aperture flat autocollimates the telescope. The WFS, NUV CAM detector, or MOS detector senses the return image. This provides a full system optical image quality test of the OTA, TFA, NUV CAM, and MOS. The PSS modes are tested at the PSS assembly level with a f/5 small field OTA simulator.

The CETUS optical alignment achieved before launch will change due to exposure to launch vibration levels and after launch to the temperature environments, both temperature levels and gradients. The design includes systems to measure the change in wavefront error and to partially correct the measured error by realignment of the OTA by SM position adjustment. In addition, tip/tilt/focus mechanisms in both the MOS and the CAM, which allow dithering the image to sample FPA pixel/resels nonuniformities, also allow realignment of these SIs using the output of their detectors as error sensors. The OTA and all instruments are designed to retain alignment by using low-thermal expansion materials for structure and optics, stable designs for the mirror mounts, and temperature control. Therefore, we anticipate that on-orbit realignment will only be required infrequently.

In summary, the CETUS optical design presented in this paper fulfills all of the requirements necessary for the CETUS science program to be successfully completed, if the CETUS probe mission is selected and advances beyond the current study phase which is nearing completion.

## Acknowledgments

The authors acknowledge the support through a NASA grant and through NASA Goddard Space Flight Center Internal Research and Development funds. RAW received subaward funding from the master NASA Grant, with Grant and Cooperative Agreement No. NNX17AL36G.

## References

1. S. Heap et al., "CETUS: an innovative UV probe-class mission concept," *Proc. SPIE* 4 (2017).
2. S. Heap et al., "The NASA probe-class mission concept, CETUS (cosmic evolution through ultraviolet spectroscopy)," *Proc. SPIE* **10398**, 103980U (2017).
3. L. Purves, "Mission systems engineering for the cosmic evolution through UV spectroscopy (CETUS) space telescope concept," *Proc. SPIE* 10401 (2017).
4. A. Hull et al., "The CETUS probe mission concept 1.5 m optical telescope assembly: a high A-Omega approach for ultraviolet astrophysics," *Proc. SPIE* 10699 (2018).
5. R. Woodruff et al., "Optical design for CETUS: a wide-field 1.5 m aperture UV payload being studies for a NASA probe class mission study," *Proc. SPIE* **10401**, 104011P (2017).
6. R. Woodruff et al., "Optical design for CETUS: a wide-field 1.5m aperture UV payload being studies for a NASA probe class mission study," in *AAS 140.15* (2018).

7. M. Bottema and R. Woodruff, "Third order aberrations in Cassegrain-type telescopes and coma correction in servo-stabilized images," *Appl. Opt.* **10**(2), 300 (1971).
8. S. Kendrick et al., "Multiplexing in astrophysics with a UV multi-object spectrometer on CETUS, a probe-class mission study," *Proc. SPIE* **10401**, 1040111 (2017).
9. S. Kendrick et al., "UV spectroscopy with the CETUS ultraviolet multi-object spectrometer (MOS)," in *AAS 140.08* (2018).
10. S. Kendrick et al., "UV capabilities of the CETUS multi-object spectrometer and NUV/FUV camera," *Proc. SPIE* **10699**, 1069939 (2018).
11. D. Korsch, *Reflective Optics*, Academic Press, Inc., San Diego, California, pp. 217–218 (1991)
12. U. Brauneck et al., "Theoretical study of filter design for UV-bandpass filters for the CETUS probe mission study," *Proc. SPIE* **10699**, 106993D (2018).
13. M. A. Quijada, J. Del Hoyo, and S. Rice, "Enhanced far-ultraviolet reflectance of MgF2 and LiF over-coated Al mirrors," *Proc. SPIE* **9144**, 91444G (2014).
14. B. Fleming et al., "Advanced environmentally resistant lithium fluoride mirror coatings for the next generation of broadband space observatories," *Appl. Opt.* **56**(36), 9941–9950 (2017)
15. A. Maréchal, "André Étude des effets combinés de la diffraction et des aberrations géométriques sur l'image d'un point lumineux," *Rev. Optique* **26**, 257–277 (1947).

**Robert A. Woodruff** has over 50 years of experience in designing optical systems for U.S. space program missions. He has made significant contributions to projects including Skylab, Nimbus, Apollo-Soyuz, Galileo, SIRTF/Spitzer, microgravity, Hubble Space Telescope, Next Generation Space Telescope (aka JWST), Terrestrial Planet Finder, Beyond Einstein, Exo-planet detection, and Kepler. He has served in various technical roles in optical design, system engineering, system test, and system calibration on more than 20 flight hardware telescopes/instruments.

**Tony Hull** is an adjunct professor of physics and astronomy at the University of New Mexico and consults extensively in the international space-optomechanics community, having practiced these disciplines for 40 years. He has served as director of Large Optics for L-3 IOS, with cognizance and management of the OpFab of the entire suite of JWST mirrors. Prior to that he was NASA's Technologist for TPF-C, principal engineer at JPL, main lead of Team-I, and founding manager of HCIT. He has served as vice president, director of Engineering and chief scientist at Optical Corporation of America, and director at Perkin-Elmer. He chairs this summer's SPIE Conference on Astronomical Telescopes, and is giving an invited talk at OSA's summer meeting. On CETUS, he is payload architect.

**Stephen E. Kendrick** has 44 years of optical and electro-optical systems experience and 33 years of technical and program management experience. He has led several technology studies and hardware demonstration programs and has been the proposal lead for several new business pursuits. His experience spans a broad range of disciplines including working on star sensors/guiders, adaptive optics, spacecraft buses, and on spaceborne optics from the Hubble Space Telescope to the present James Webb Space Telescope.

Biographies of the other authors are not available.